\documentclass{aa}  

\usepackage{graphicx}
\usepackage{subfig}
\usepackage{xcolor}
\usepackage{amsmath}
\usepackage{amssymb}
\usepackage{mathtools}
\usepackage{amsfonts}
\usepackage{longtable}
\usepackage{natbib}
\usepackage[percent]{overpic}
\usepackage{tikz}
\usetikzlibrary{arrows.meta, positioning, shapes.geometric, fit, backgrounds, calc}
\usepackage{booktabs}
\usepackage{adjustbox}
\usepackage[T1]{fontenc}
\usepackage{newtxtext,newtxmath}

\usepackage{multirow}
\usepackage[normalem]{ulem}
{\bfseries}{\itshape}
\usepackage{float}
\usepackage{enumitem}

\setcitestyle{authoryear,round,aysep={,},yysep={,},date=year}

\makeatletter
\renewcommand{\fnum@figure}{\textbf{Fig.\ \thefigure}}
\makeatother

\makeatletter
\renewcommand{\fnum@table}{\textbf{Table\ \thetable}}
\makeatother

\usepackage{indentfirst}
\usepackage{hyperref}

\hypersetup{colorlinks=true,linkcolor=blue,citecolor=blue,filecolor=blue,urlcolor=blue}

\makeatletter
\@fleqnfalse
\makeatother

\usepackage{placeins}

\titlerunning{{\mdseries \textsc{Chronos:}} I. A Bayesian hierarchical lithium-age model: Validation on the Pleiades cluster}

\begin{document}

   \title{{\mdseries \textsc{Chronos:}} Towards a self-consistent and absolute stellar age scale}
   \subtitle{I. A Bayesian hierarchical lithium-age model: Validation on the Pleiades cluster}
    
   \author{L. González-Ramírez\inst{1,2}\thanks{Corresponding author: \href{mailto:lgrjr27798@gmail.com}{lgrjr27798@gmail.com}}
   \and D. Barrado\inst{1}
   \and J. Olivares \inst{3}
   \and A. Berihuete\inst{4}
   \and L. M. Sarro\inst{3}
   \and F. J. Palmero\inst{5}
   }
   
   \institute{
        Centro de Astrobiolog\'ia (CAB), CSIC-INTA, Camino Bajo del Castillo s/n, E-28692 Villanueva de la Ca\~nada, Madrid, Spain
        \and
        Departamento de Física de la Tierra y Astrof\'isica, Facultad de Ciencias F\'isicas, Universidad Complutense de Madrid, 28040 Madrid, Spain
        \and
        Departamento de Inteligencia Artificial, Universidad Nacional de Educaci\'on a Distancia (UNED), c/Juan del Rosal 16, E-28040, Madrid, Spain. 
        \and 
        Dpto. Estad\'istica e Investigaci\'on Operativa, Universidad de C\'adiz, Avda. Rep\'ublica Saharaui s/n, 11510 Puerto Real, C\'adiz, Spain.
    \and
        Departamento de \'Algebra, Geometr\'ia y Topolog\'ia, Universidad de M\'alaga, Ap. 59, 29080 M\'alaga, Spain.
    }

    \date{Received , - ; accepted }
 
\abstract
{Establishing a self-consistent age scale for stellar populations requires physically well-calibrated chronometers. Among these, lithium-based diagnostics, particularly the lithium depletion boundary (LDB), provide one of the most robust age constraints for stellar populations in the low-mass regime. However, their application is limited by heterogeneous temperature scales and astrophysical dispersion, especially among FGK stars, where rotation can significantly affect lithium abundances.}
{As a first step towards a self-consistent age scale, our aim has been to formulate \textsc{Chronos}, the first version of a Bayesian hierarchical lithium-based age-dating model combined with a neural network trained on stellar evolutionary models.}
{We implemented a Bayesian hierarchical model that jointly infers stellar effective temperatures, lithium abundances, and the global age of a stellar association. The theoretical LDB is provided by a pre-trained multilayer perceptron based on BT--Settl evolutionary models. The model incorporates a temperature-dependent transition between fully convective ultra-cool dwarfs (UCDs) and FGKM dwarf stars, together with a two-component FGK mixture to account for rotation-induced lithium enhancement. We applied the method to the Pleiades cluster and performed a validation using synthetic datasets.}
{For the Pleiades cluster, \textsc{Chronos} yields a posterior age distribution centred at $\mathrm{Age}=124.53_{-2.70}^{+3.34}$\,Myr, consistent with classical LDB estimates, while simultaneously constraining both global and stellar-level rotation parameters.}
{This work demonstrates that lithium-based stellar chronology can be recast as a coherent hierarchical inference problem, providing a flexible and statistically robust framework for making age determinations for young (1--600 Myr) stellar populations.}


\maketitle
\nolinenumbers

\section{Introduction}\label{sc:intro}

\noindent Despite the advent of stellar chronology in the early 20th century, making accurate determinations of stellar ages remains a central challenge in astrophysics. Significant progress in the systematic dating of stars occurred in the 1960s, when some of the first observational age indicators for solar-type stars were introduced \citep[e.g.][]{Wilson1963, Skumanich1972}.

Within this broad context, the lack of systematic cross-calibration between different stellar dating techniques (often yielding discrepancies of up to a factor of 2) highlights the need for a unified and physically anchored age framework. Together, two anchors provide the only absolute reference points available for tying stellar ages across the full cosmic time domain. The first anchor is the age of the Universe, estimated at $13.799 \pm 0.021$ Gyr from cosmological observations obtained by the Planck mission within the theoretical $\Lambda$cold dark matter ($\Lambda$CDM) model \citep{Planck2015}. The second is the age of the Solar System and, by extension the Sun, both of which formed from the same collapsing protostellar cloud.

The age of the Solar System is constrained through highly precise and model-independent radiometric dating of primitive Solar System materials \citep[$4.55$ Gyr,][]{Patterson1956}. This remains the only case in which an astrophysical age can be determined in a truly direct way, relying on nuclear decay as a physical clock. Complementary estimates based on stellar evolution models and helioseismology yield consistent values \citep[e.g.][]{Bonanno2002,Amelin2002,Baker2005}.

Among current stellar age-dating techniques, which span a wide range of physical assumptions and model dependencies, the lithium depletion boundary (LDB) method \citep{Rebolo1992,Magazzu1993} is a semi-fundamental chronometer directly linking stellar physics to age \citep{Soderblom2014}. For this reason, we adopted the LDB technique as the first building block of the \textsc{Chronos} framework. In recent years, LDB ages have been successfully derived for several well-studied clusters and young associations, including the Pleiades, Alpha Persei, IC 2391, and the Hyades \citep[e.g.][]{Basri1996,Stauffer1998,Barrado2004,Martin2018}, as well as the $\beta$ Pictoris moving group, in agreement with isochrone fitting \citep[$20 \pm 10$ Myr;][]{Barrado1999}.

From a physical standpoint, the LDB marks the point in a coeval stellar population at which lithium is depleted in low-mass stars, producing a sharp and age-sensitive feature observable through spectroscopy. Lithium destruction occurs once pre-main sequence stars contract sufficiently for their central temperatures to reach $\sim3\times10^{6}$ K, triggering proton-induced burning of $^{7}$Li \citep[e.g.][]{Basri1996,Chabrier1996,Bildsten1997,Ushomirsky1998}. In fully convective low-mass stars, lithium is rapidly depleted, producing a sharp transition in lithium abundance at a characteristic effective temperature for a given age, whereas warmer stars progressively develop radiative cores and exhibit more complex lithium behaviour influenced by additional processes such as rotation \citep[e.g.][]{Chabrier1996,Chabrier1997,Bildsten1997}.

\begin{figure}[!t]
\centering
\includegraphics[width=\columnwidth, trim=0.25cm 1.85cm 0.25cm 1.80cm, clip]{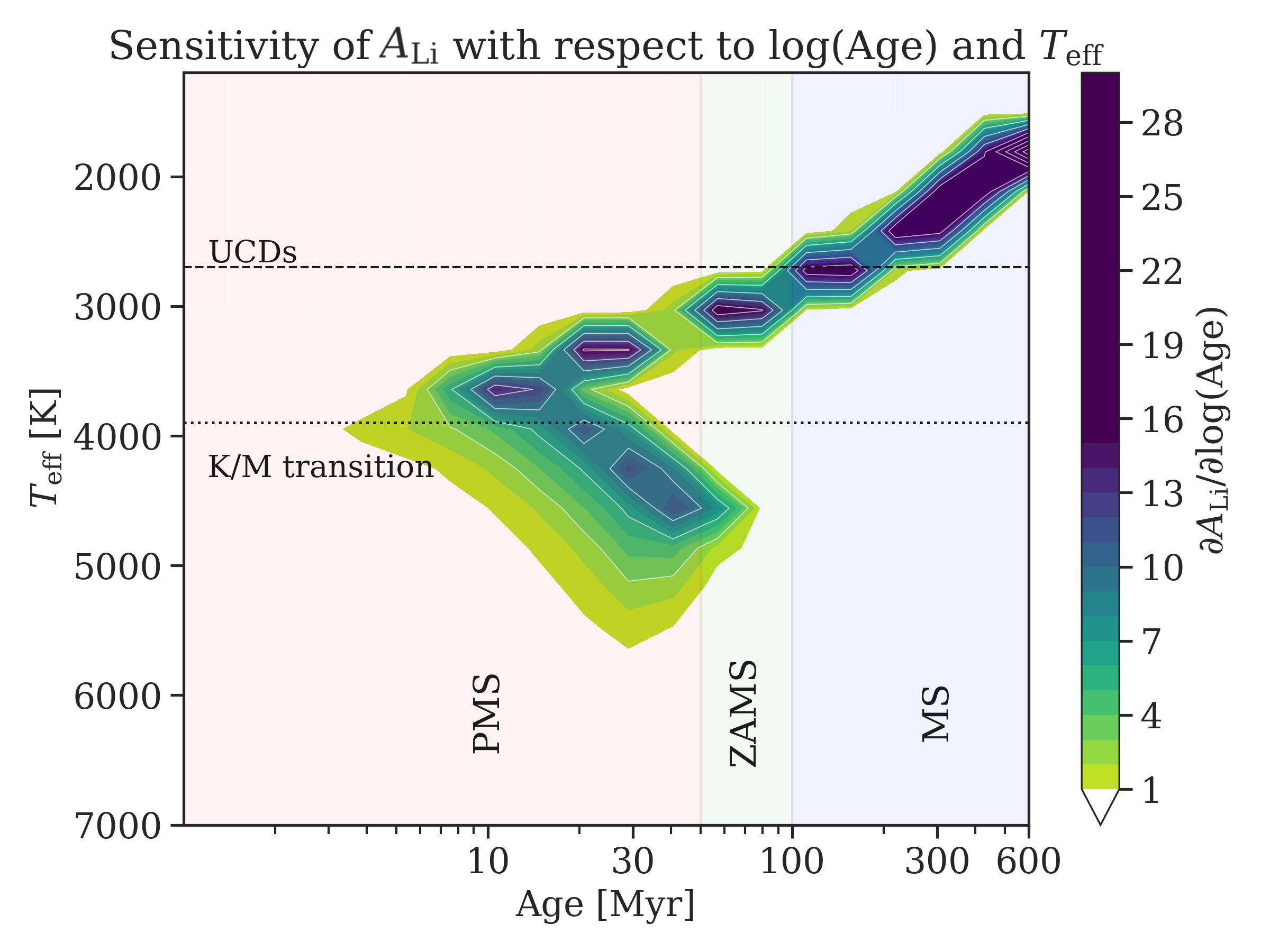}
\caption{2D sensitivity map of $A_{\rm Li}$ to stellar age, $\partial A_{\rm Li}/\partial \log{(\mathrm{Age})}$, as a function of $T_{\rm eff}$. Age sensitivity is highest in the fully convective regime at low $T_{\rm eff}$, while the sensitivity decreases towards warmer FGK stars. The apparent clumpy structure, visible for $T_{\rm eff} > 4000$ K, is a numerical artefact arising from the discrete age sampling of the BT-Settl grid.}
\label{fig:sens}
\end{figure}

Consequently, fully convective stars display a sharp and well-defined LDB due to efficient mixing, while partially radiative FGK stars follow a more complex lithium evolution shaped by additional processes such as rotation and magnetic activity. The transition between the two regimes occurs over a narrow but finite range of effective temperatures around $T_0 \simeq 3700\,\mathrm{K}$ and, therefore, it has a limited impact on age inference, as the sensitivity of lithium depletion to age peaks in the fully convective regime and decreases rapidly at higher temperatures. This behaviour is illustrated in Fig. \ref{fig:sens}, which shows a two-dimensional map of the sensitivity of the lithium abundance (hereafter $A_{\rm Li}$) to age, $\partial A_{\rm Li}/\partial \log{(\mathrm{Age})}$, as a function of effective temperature (hereafter $T_{\rm eff}$). We note that the age sensitivity is maximised at low $T_{\rm eff}$. For this reason, using a $T_{\rm eff}$ range extending to ultra-cool dwarfs (UCDs) is essential for LDB age inference (including brown dwarfs).

Even with its empirical success, most existing LDB-based age determinations rely on sharp temperature or luminosity cuts to isolate stellar objects assumed to belong to the fully convective regime, together with single-valued prescriptions that map a given observable to a unique age. While these choices are often necessary in practice, they implicitly assume an abrupt regime transition and prevent a consistent propagation of observational and modelling uncertainties across the transition region.

\begin{figure}[!t]
\includegraphics[width=\columnwidth, trim=0.25cm 0.35cm 0.25cm 1.425cm, clip]{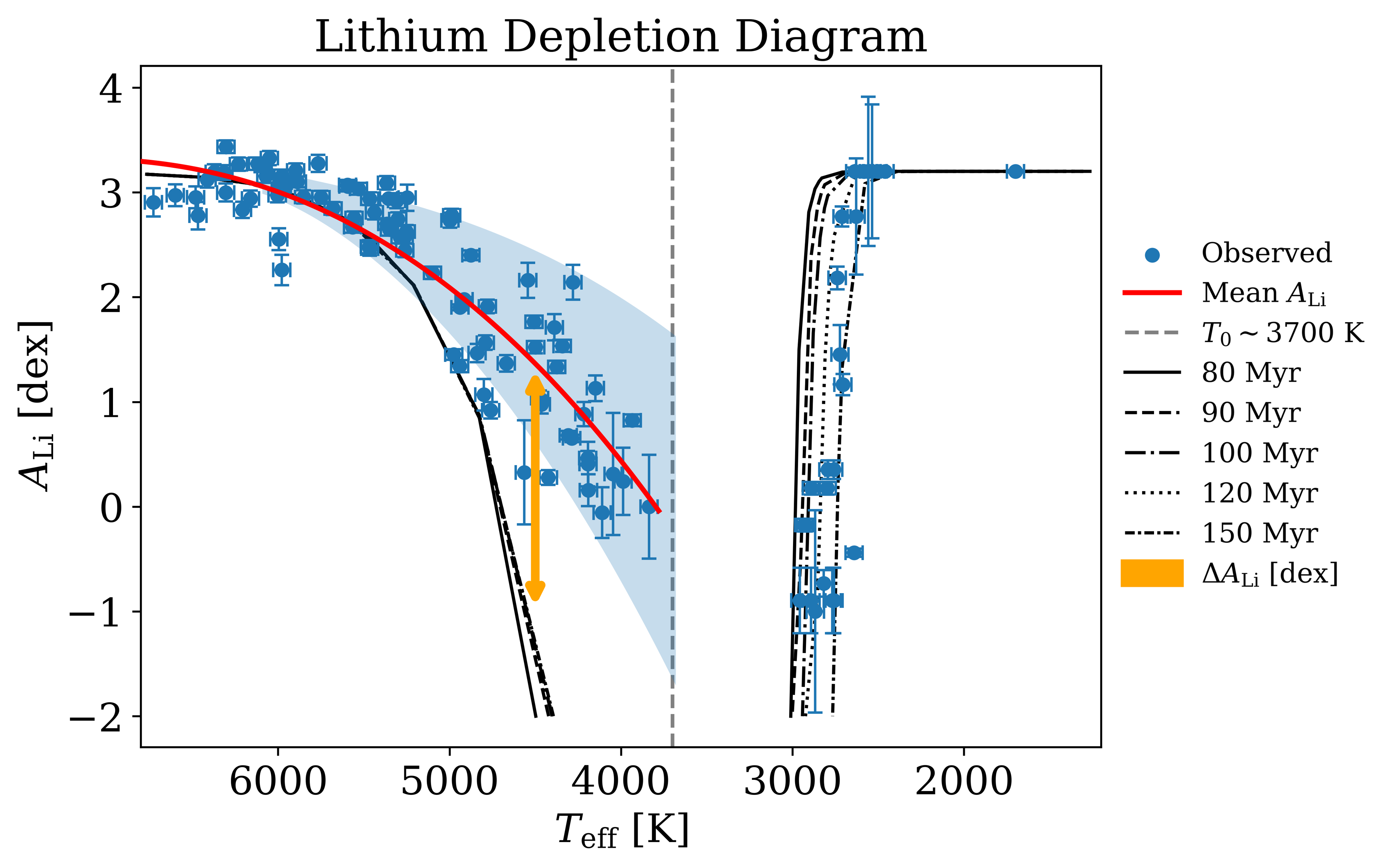}
\caption{Lithium depletion diagram for the Pleiades, using the dataset described in Sect. \ref{sc:data}. Blue points show individual stellar $A_{\rm Li}$ with associated uncertainties. The solid red curve represents a smooth estimate of the mean $A_{\rm Li}$ and the vertical dashed line is at 3700 K. Black curves show BT-Settl models for several ages around 120 Myr. The approximate expected $\Delta A_{\rm Li}$ offset due to the rotation enhancement is shown. The width of this band does not correspond to a confidence interval.}
\label{fig:intro_ldd}
\end{figure}

Additional complexity arises within the FGK stellar population itself. Observations indicate that young FGK stars exhibit a broad and asymmetric lithium distribution at fixed effective temperature and age, strongly correlated with stellar rotation \citep[e.g.][]{Barrado2016,Bouvier2018}. Fast rotators systematically display enhanced $A_{\rm Li}$ and increased dispersion relative to their slowly rotating counterparts. While the transition between slow and fast rotators is not strictly discrete, the resulting lithium enhancement is sufficiently pronounced to require explicit modelling of this population. Quantitatively, individual rapidly rotating stars can exhibit $A_{\rm Li}$ values up to a factor of $\sim2$ higher than slow rotators at ages of a few million years (Myr). By the zero-age main sequence, as in the Pleiades, this enhancement can reach values of the order of $600\%$ \citep{Jeffries2021,Jeffries2023}. These differences correspond to logarithmic abundance offsets up to $\Delta A_{\rm Li}\simeq 3.0$ dex. This interpretation is further supported by observational studies of Li\,\textsc{I} equivalent width (EW) as a function of rotation period, in which the rapidly rotating sequence is systematically offset towards larger EW values relative to the slow-rotator sequence, particularly among K-type stars \citep{Cuenda2025}. This behaviour is directly reflected in the observed $A_{\rm Li}$ and illustrated schematically in the lithium depletion diagram (Fig. \ref{fig:intro_ldd}), where the shaded band provides an approximate guide to the range of lithium enhancement and increased dispersion associated with stellar rotation at fixed age.

Recent studies have shown that this asymmetric lithium dispersion emerges early in the pre-main sequence phase and is already present at ages of $\sim10$--$20$ Myr, coincident with the onset of lithium burning \citep{Jackson2025}. These results challenge standard evolutionary descriptions in which lithium depletion is treated as a single-valued function of mass and age, and motivate the use of statistical frameworks capable of accommodating intrinsic astrophysical diversity.

Given the observational limitations associated with lithium-based age diagnostics, particularly for low-mass stars, robust statistical modelling is required to extract reliable age constraints. Bayesian hierarchical models (BHMs; see \citealt{Gelman2013} for a detailed review) provide a natural framework for modelling complex and multi-level data structures, while consistently propagating uncertainties.

Bayesian hierarchical models have also been successfully applied in stellar astrophysics, including hierarchical treatments of stellar clusters and populations \citep[e.g.][]{daSilva2006,Rodrigues2014,Sarro2014,Stenning2016,Si2018,Valcin2025,Lyttle2021,Mas-Buitrago2024,Saunders2024}. In particular, \citet{Olivares2018} applied a BHM to nearby young clusters within the DANCe project \citep[Dynamical Analysis of Nearby Clusters;][]{Bouy2013}, from which the Pleiades sample used here is mostly drawn. A further application to stellar ages is presented in works such as \citet{Moore2020} and \citet{Jeffries2023b} (\textsc{eagles} model), both of which adopt an empirical approach based on Li\,\textsc{I} EW. Collectively, these studies illustrate the strength of statistical frameworks for combining heterogeneous observables and uncertainties, while also revealing limitations when complex non-linear lithium depletion is represented through simplified or purely empirical prescriptions.

In parallel, artificial neural networks (ANNs) have proven effective in modelling complex, high-dimensional astrophysical data. Neural networks have been extensively employed for the emulation of stellar models and the interpolation of high-dimensional evolutionary grids \citep[e.g.][]{vonHippel1994, Maltsev2024,Hon2024,Teng2025,Scutt2026}, providing fast and flexible surrogates that preserve the underlying physical trends. In this context, ANNs provide an alternative to discrete evolutionary grids, formally supported by the universal approximation theorem\footnote{The universal approximation theorem ensures that a feed-forward neural network with at least one hidden layer containing a finite number of neurons with non-linear activations can approximate any continuous function on a compact subset of $\mathbb{R}^{n}$ to arbitrary precision.}. More recently, along similar lines, \citet{Weaver2024} extended \textsc{eagles} by introducing an ANN trained on the same dataset as \citet{Jeffries2023b}, achieving a substantially improved representation of Li\,\textsc{I} EWs and their intrinsic dispersion. This work demonstrates the advantages of ANN-based approaches over purely analytical prescriptions in capturing complex lithium patterns. At the same time, it shows that increased flexibility alone is not sufficient to fully capture lithium depletion behaviour across different stellar populations, particularly at ages $\gtrsim 1$ Gyr.

Motivated by these developments, we present a BHM for lithium-based age determination. The model jointly infers cluster age, as well as the $T_{\rm eff}$ and $A_{\rm Li}$ of its members, while accounting for intrinsic dispersion within a coherent probabilistic framework. More broadly, \textsc{Chronos} represents a first step towards a unified framework in which multiple age diagnostics can be incorporated within a single hierarchical model, allowing their relative constraining power to be assessed self-consistently. The approach naturally captures regime transitions and rotation-related behaviour without imposing sharp temperature thresholds. Using synthetic data and the Pleiades as a benchmark, we demonstrate that the model robustly recovers a well-established cluster age.

The aim of this work is to assess the reliability of the code and to refine its applicability to age-dating young stellar populations using the LDB technique. This paper is organised as follows. Section 2 outlines the model architecture and the methodology applied to both synthetic and real observations. In Sect. 3 we describe the dataset. In Sect. 4 we present a model sensitivity analysis to justify the model architecture. In Sect. 5 we present our validation results and compare them with previous studies. Finally, Sects. 6 and 7 provide our discussion and conclusions, along with future perspectives.

\section{Methods}

\noindent Bayesian hierarchical modelling provides a natural framework to coherently combine heterogeneous sources of information and to explicitly propagate uncertainties across different hierarchical levels \citep{Gelman2013}. This approach is particularly advantageous when dealing with mixed spectroscopic datasets, which often differ in terms of precision and completeness.

\subsection{Bayesian hierarchical model}\label{sc:bhm}

\noindent The goal of \textsc{Chronos} is to infer the age of a stellar cluster based on $T_{\rm eff}$ and $A_{\rm Li}$, while consistently propagating observational uncertainties and accounting for population-level effects that can bias lithium-based age estimates, such as intrinsic dispersion and stellar rotation. The statistical structure of the model is summarised schematically in Fig. \ref{fig:tikz_model}, and is described below.

The Bayesian hierarchical model adopted in this work follows directly from Bayes’ theorem. With $\Theta$ denoting the full set of global parameters (explained in Sect. \ref{sc:bhm:global}), the joint posterior distribution of the hierarchical model is then given by
\begin{equation}\label{eq:bayes}
p(\Theta,\boldsymbol{Z}\mid\boldsymbol{D})
=
\frac{
p(\boldsymbol{D}\mid\boldsymbol{Z})\,
p(\boldsymbol{Z}\mid\Theta)\,
p(\Theta)}
{
p(\boldsymbol{D})
}
,\end{equation}
where $\boldsymbol{Z}$ denotes the set of latent variables, with its joint distribution $p(\boldsymbol{Z}\mid\Theta)$ detailed in Sect. \ref{sc:bhm:prior}, and $\boldsymbol{D}$ the observed data, with their reported uncertainties, $\sigma_{{T_{\rm eff},\rm obs},i}$ and $\sigma_{{A_{\rm Li},\rm obs},i}$, via
\[
\boldsymbol{D} =
\left\{
T_{{\rm eff,obs},i},\,
A_{{\rm Li,obs},i}
\right\}_{i=1}^{N},
\]
where $i=1,\dots,N$ indexes the sources in the sample.

The strategy adopted here is intentionally non-prescriptive. Rather than pre-assigning observations to discrete classes or enforcing sharp boundaries (e.g. slow versus fast rotators, or fully convective versus partially radiative sources defined by a hard $T_{\rm eff}$ threshold), the model embeds these physical considerations within a hierarchical Bayesian framework. In particular, spectral regime and rotation effects on $A_{\rm Li}$ are treated probabilistically.

\subsubsection{Global parameters}\label{sc:bhm:global}

\noindent At the highest level of the hierarchy, the model includes a set of global parameters $\Theta = \{\mathrm{Age},\, \sigma_{A_{\rm Li}},\, \theta_{\rm rot},\, \theta_r\}$, which describe the properties of the stellar population as a whole. The cluster age is the primary parameter of interest. The term $\sigma_{A_{\rm Li}}$ represents an intrinsic lithium-abundance dispersion added to the theoretical models (noiseless) predictions, capturing astrophysical variance not explained by the model.

Rotation-related effects in the FGK regime are described (see Sect. \ref{sc:mixture} for a detailed formulation) by the parameter set $\theta_{\rm rot} = \big\{ \Delta A_{{\rm Li},\rm rot},\, \sigma_{A_{\rm Li},\rm rot},\, p_{\rm rot},\, \alpha_{\rm FGK} \big\}$, where $\Delta A_{{\rm Li},\rm rot}$ controls the amplitude of the lithium enhancement associated with rotation, $\sigma_{A_{\rm Li},\rm rot}$ the additional dispersion among the fast rotators, $p_{\rm rot}$ the fraction of stars with rotation-induced lithium enhancement relative to the theoretical abundances and $\alpha_{\rm FGK}$ the skewness of the lithium distribution in this population.

For reasons explained in the introduction (Sect. \ref{sc:intro}), the distribution of $A_{\rm Li}$ for a given $T_{\rm eff}$ is modelled differently for fully convective and partially radiative sources (including rotation-induced lithium enhancement). The transition between fully convective stellar objects and partially radiative FGK stars is described by $\theta_r = \{T_0,\, \sigma_{T_0}\}$, where $T_0$ denotes the characteristic transition temperature and $\sigma_{T_0}$ its width.

\subsubsection{Local parameters}\label{sc:bhm:local}

\noindent At the lowest hierarchical level, each star $i$ is characterised by a set of local parameters described by latent variables,
\[
\boldsymbol{Z}_i = \{T_{{\rm eff,true},i},\, A_{{\rm Li,true},i}\},
\]
representing the true $T_{\rm eff}$ and $A_{\rm Li}$. These quantities are not observed directly. In particular, $T_{{\rm eff,true},i}$ is assigned an individual prior distribution $p(T_{{\rm eff,true},i})$, while $A_{{\rm Li,true},i}$ is treated as a latent stochastic quantity drawn from a population-level conditional distribution $p(A_{{\rm Li,true},i}\mid T_{{\rm eff,true},i},\Theta)$, parameterised by $\Theta$ and $T_{{\rm eff,true},i}$ (see Sect. \ref{sc:mixture}, Eq. \ref{eq:latent_two_component_mixture}).

The probability that a star belongs to the FGK regime, $\omega_{\mathrm{FGK},i}$, is modelled as a smooth function of its true effective temperature,
\begin{equation}
\omega_{\mathrm{FGK},i}
=
\mathrm{sigmoid}\!\left[\left(
T_{\mathrm{eff,true},i}-T_0\right)/\sigma_{T_0}\right]
\label{eq:sigmoid},
\end{equation}
ensuring a continuous transition between fully convective and partially radiative objects.

\subsubsection{Rotation mixture}\label{sc:mixture}

\noindent Then, the prior probability that star $i$ belongs to the population of fast rotators is defined as $p(\mathrm{rot}_i)$. This probability is defined as a marginalisation over the spectral regime:
\begin{equation}
\begin{split}
    p(\mathrm{rot}_i) & = 
    p(\mathrm{rot}_i \mid \mathrm{FGK}_i)\,\times\,p(\mathrm{FGK}_i) \\
    & + p(\mathrm{rot}_i \mid \overline{\mathrm{FGK}_i})\,\times\,p(\overline{\mathrm{FGK}_i}),
\end{split}
\end{equation}
where $1 - \omega_{\mathrm{FGK},i}= p(\overline{\mathrm{FGK}_i})$ represents the probability that the star is fully convective.

Since rotational lithium enhancement is expected to be negligible in the fully convective regime \citep[because these objects lack the tachocline required for solar-type dynamo action and rotation-dependent structural inflation; see][]{Chabrier1997,Reiners2008,Somers2015}, we set $p(\mathrm{rot}_i \mid \overline{\mathrm{FGK}_i})=0$.

Under this assumption, the probability that star $i$ belongs to the population of fast rotators is reduced to
\begin{equation} \label{eq:prob_i}
    \omega_i
    =
    p(\mathrm{rot}_i \mid \mathrm{FGK}_i)\,\times\,p(\mathrm{FGK}_i)
    =
    p_{\rm rot}\,\times\,\omega_{\mathrm{FGK},i},
\end{equation}
where $p_{\rm rot} \equiv p(\mathrm{rot}_i \mid \mathrm{FGK}_i)$ is a global parameter representing the fraction of fast rotators within the FGK population.

Second, conditionally on the cluster age and the true stellar effective temperature, we can use a multilayer perceptron (NN; see Sect. \ref{sc:mlp} for details) to predict the theoretical lithium abundance for slow rotators, $\mu_{A_{\mathrm{Li}},i} \equiv f_{\mathrm{NN}}(\mathrm{Age},T_{\mathrm{eff,true},i})$, where $f_{\mathrm{NN}}$ denotes the trained NN. This quantity defines the location of the conditional distribution of $A_{{\rm Li,true},i}$, rather than the lithium abundance itself.

Then, the true lithium abundance of each star, represented by the latent variable $A_{{\rm Li, true},i}$, is modelled as a finite mixture of two populations, one representing the slow rotators and the fast rotators in the other.

We define the mixture as
\begin{equation}
\begin{split}
& p(A_{\mathrm{Li,true},i}\mid T_{{\rm eff,true},i},\,\Theta) = \,(1-\omega_i)\,\times\,
\mathcal{N}\,\left(
\mu_{A_{\mathrm{Li}},i},\,
\sigma_{A_{\rm Li}}^2
\right)
\\[4pt]
&+
\omega_i\,\times\,
\mathcal{SN}\,\left(
\mu_{A_{\mathrm{Li}},i}
+
\Delta A_{\rm Li,rot},\,
\sigma^2_{\rm tot},\,
\alpha_{\rm FGK}
\right),
\end{split}
\label{eq:latent_two_component_mixture}
\end{equation}
where $\sigma^2_{\rm tot} = \sigma_{A_{\rm Li}}^2 + \sigma_{A_{\rm Li},\rm rot}^2$ and the weight $\omega_i \in (0,1)$ represents the probability that star $i$ belongs to the fast rotating population given by Eq. (\ref{eq:prob_i}). $\mathcal{SN}$ denotes a Skew-Normal distribution (see Appendix \ref{ap:distributions}).

\begin{figure}[t!]

\begin{tikzpicture}[
    node distance=20mm and 22mm,
    latent/.style = {draw, ellipse, rounded corners=2pt, minimum width=15mm, minimum height=8mm, align=center, fill=white},
    obs/.style    = {draw, ellipse, minimum width=15mm, minimum height=5mm, align=center, fill=gray!25},
    det/.style    = {draw, rectangle, aspect=2.0, minimum width=8mm, minimum height=6mm, align=center, fill=white},
    plate/.style  = {draw, rectangle, inner sep=3.5mm, rounded corners=3pt},
    arrow/.style  = {-{Stealth[length=3mm,width=2mm]}, semithick, black},
    darrow/.style = {dashed, -{Stealth[length=3mm,width=2mm]}, semithick, black}
]

\node[det] (NN) at (+8.5,0) {$f_{\rm NN}$};
\node[det, left=of NN, xshift=0.05cm, yshift=-1.05cm] (wSpec) {$\omega_{{\rm FGK},i}$};
\node[latent, left=of NN, xshift=-0.75cm] (trueTeff) {$T_{{\rm eff, true},i}$};
\node[latent, below=of NN, yshift=+1.5cm] (trueALi) {$A_{{\rm Li,true},i}$};
\node[obs, below=of NN, yshift=+0.25cm] (obsALi) {$A_{{\rm Li,obs},i}$};
\node[obs, below=of trueTeff, yshift=+0.35cm] (obsTeff) {$T_{{\rm eff,obs},i}$};

\node[latent, below=of wSpec, yshift=+0.25cm] (theta_r) {$\theta_{r}$};
\node[latent, right=of NN, xshift=-1.cm] (Age) {$\mathrm{Age}$};
\node[latent, right=of obsALi, xshift=-1.5cm] (sigmaALi) {$\sigma_{A_{\rm Li}}$};

\node[latent, right=of trueALi, xshift=-1.5cm] (theta_rot) {$\theta_{\rm rot}$};
\node[det, right=of wSpec, xshift=-1.85cm, yshift=-1.00cm] (PstarDet) {$\omega_i$};
\node[latent, below=of wSpec, xshift=+2.25cm, yshift=+.25cm] (Pstar) {$p_{\rm rot}$};

\node[plate, fit=(wSpec)(NN)(trueALi)(PstarDet)(trueTeff)(obsALi)(obsTeff), label=above:{$i = 1,\dots,N$}, yshift=0.25cm] (plate) {};

\draw[arrow] (trueTeff) -- (wSpec);
\draw[arrow] (Pstar) -- (PstarDet);
\draw[darrow] (wSpec) -- (PstarDet);
\draw[arrow] (trueTeff) -- (obsTeff);
\draw[darrow] (PstarDet) -- (obsALi);
\draw[arrow] (trueTeff) -- (NN);
\draw[darrow] (NN) -- (trueALi);
\draw[arrow] (trueALi) -- (obsALi);

\draw[arrow] (theta_r) -- (wSpec);
\draw[arrow] (Age) -- (NN);
\draw[arrow] (sigmaALi) -- (trueALi);
\draw[arrow] (theta_rot) -- (trueALi);


\node[font=\bfseries] at ([yshift=-8mm, xshift=-4mm]current bounding box.north){\textsc{Chronos} BHM};

\end{tikzpicture}
\caption{Schematic workflow of \textsc{Chronos} BHM model. We set white ellipses for parameters and grey ellipses for observations, rectangles for deterministic nodes, and continuous arrows for conditional relations and dashed for deterministic relation.}
\label{fig:tikz_model}
\end{figure}

\subsubsection{Likelihood}

\noindent Observed quantities are linked to the latent stellar parameters by likelihoods representing Gaussian measurement models. For each source $i$, we assume conditional independence between the $T_{\rm eff}$ and $A_{\rm Li}$ measurements given the latent variables, so that the global likelihood $\mathcal{L} \equiv p(\boldsymbol{D} \mid \boldsymbol{Z})$ is represented by the product of individual source-by-source likelihoods,
\begin{equation}
\begin{split}
\prod^{N}_{i=1}\, \mathcal{L}^{(i)}_{A_{\rm Li}}\times \mathcal{L}^{(i)}_{T_{\rm eff}} = & 
\prod_{i=1}^{N}
\mathcal{N}\!\left(A_{\mathrm{Li,obs},i}\mid A_{\mathrm{Li,true},i},\sigma_{{A_{\rm Li},\rm obs},i}\right)\\
& \, \times \,
\mathcal{N}\!\left(T_{\mathrm{eff,obs},i}\mid T_{\mathrm{eff,true},i},\sigma_{{T_{\rm eff},\rm obs},i}\right)
\label{eq:likelihood},
\end{split}
\end{equation}

We emphasise that the assumption of conditional independence between $A_{\rm Li}$ and $T_{\rm eff}$ constitutes an approximation. In practice, $A_{\rm Li}$ is derived from the measured EW through $T_{\rm eff}$-dependent curves of growth (CoGs) such that part of its uncertainty is formally correlated with that of $T_{\rm eff}$.

A fully self-consistent hierarchical formulation would require modelling lithium EW directly, introducing latent $\mathrm{EW}_{\rm true}$ and embedding the curve-of-growth transformation $A_{\rm Li}=g(\mathrm{EW},T_{\rm eff})$ within the probabilistic framework, thereby naturally propagating this dependence and the associated covariance structure at the likelihood level. In the present work, we instead operate in $A_{\rm Li}$ space, as the evolutionary models provide lithium fractions that are deterministically mapped to $A_{\rm Li}$, and we treat the reported $A_{\rm Li}$ and $T_{\rm eff}$ uncertainties as independent, as a pragmatic approximation when combining heterogeneous literature measurements (see Sect. \ref{sc:data}). Embedding EW directly within the model constitutes a natural extension of this work. This approximation is further alleviated by the stochastic structure of the model: although the NN provides a $T_{\rm eff}$-dependent mean relation, $A_{{\rm Li,true},i}$ is drawn from a mixture distribution that includes intrinsic dispersion and rotation-induced offsets. This broadens $p(A_{{\rm Li,true},i}\mid T_{{\rm eff,true},i},\Theta)$ and reduces the sensitivity of the likelihood to the exact covariance between $A_{\rm Li}$ and $T_{\rm eff}$.

Although the model is primarily aimed at inferring the cluster age, it also yields posterior distributions for the parameters governing rotation-induced lithium enhancement, the FGK–UCD transition, and the latent variables $T_{{\rm eff,true},i}$ and $A_{{\rm Li,true},i}$, ensuring self-consistent uncertainty propagation across the different levels of the model.

\subsubsection{Priors}\label{sc:bhm:prior}

\noindent All prior distributions adopted in \textsc{Chronos} are physically motivated when appropriate and sufficiently broad, so that posterior inference is primarily driven by the data when informative. The priors are grouped according to the hierarchical levels of the model, distinguishing between global population parameters and local (source-by-source) parameters.

Since $T_{{\rm eff,true},i}$ does not depend on $\Theta$, the joint distribution of the latent variables (middle term in Eq. \ref{eq:bayes}) can be factorised, for each source, into an independent prior on $T_{{\rm eff,true},i}$ and a conditional distribution for $A_{{\rm Li,true},i}$ (see Eq. \ref{eq:latent_two_component_mixture}) expressed as
\begin{equation}\label{eq:latent_p}
p(\boldsymbol{Z}\mid\Theta)
=
\prod_{i=1}^{N}
p(T_{{\rm eff,true},i})\,
p(A_{{\rm Li,true},i}\mid T_{{\rm eff,true},i},\Theta),
\end{equation}
where $p(T_{{\rm eff,true},i})$ is the prior on the true effective temperature and Eq. \ref{eq:latent_two_component_mixture} provides the definition of the last term. This factorisation makes it explicit that $A_{{\rm Li,true},i}$ is not a deterministic function of the model parameters, but a latent stochastic quantity whose distribution is governed by $\Theta$. The prior on $T_{{\rm eff,true},i}$ is weakly informative and spans the full range of temperatures covered by the stellar models and observations.

We adopted a uniform prior for the cluster age ensuring that the posterior age constraints arise from the LDB itself. In particular, we set an age range of 10--500 Myr, where the BT-Settl models are most effective for the LDB method. For the intrinsic lithium dispersion, we used a Gamma prior to ensure positivity and to regularise unrealistically large uncertainties. We parametrised the Gamma distribution with the shape, $\alpha$, and rate, $\beta$.

The transition between the FGK and UCD/M regimes was modelled through a smooth logistic function of effective temperature. The corresponding parameters, $T_0$ and $\sigma_{T_0}$, were assigned informative yet flexible priors, centred on the theoretical boundary for full convection, with sufficient width to allow the data to constrain the transition (see Appendix \ref{ap:w_spec} for a justification of the distribution parameters $T_0$ and $\sigma_{T_0}$, included in Table \ref{tab:chronos_priors}).

Finally, the rotation-related parameters governing the fraction of FGK fast rotators and the amplitude, dispersion, and asymmetry of the rotation-induced lithium enhancement were assigned weakly informative priors that regularise the inference towards physically plausible values without severely constraining the posterior in the data-dominated regime. A complete summary of all prior distributions is provided in Table \ref{tab:chronos_priors}.

\begin{table}[t!]
\small
\centering
\caption{Prior distributions adopted.}
\begin{tabular}{ll@{\hspace{14.5pt}}c@{\hspace{12.5pt}}c@{\hspace{12.5pt}}l}
\toprule
\textbf{Parameter} & \textbf{Distribution} & \textbf{Mode} & \textbf{$\sigma$} & \textbf{Units} \\[4pt]
\midrule

\multicolumn{5}{l}{\textbf{Local parameters}} \\[4pt]
\midrule

$T_{{\rm eff,true},i}$ 
& $\mathcal{U}(1200,\;7000)$ 
& -- 
& -- 
& K \\[2pt]

\midrule

\multicolumn{5}{l}{\textbf{Main global parameters}} \\[4pt]
\midrule

$\mathrm{Age}$ 
& $\mathcal{U}(10,\;500)$ 
& -- 
& --
& Myr \\[2pt]

$\sigma_{A_{\rm Li}}$ 
& $\Gamma(\alpha=10,\;\beta=30)$ 
& $0.300$ 
& $0.105$ 
& dex \\[2pt]

\midrule

\multicolumn{5}{l}{\textbf{Global parameters for regime transition} $\theta_r$} \\[4pt]
\midrule

$T_0$ 
& $\mathcal{N}(\mu=3700,\;\sigma=25)$ 
& $3700$ 
& $25$ 
& K \\[2pt]

$\sigma_{T_0}$ 
& $\Gamma(\alpha=100,\;\beta=1)$ 
& $99$ 
& $10$ 
& K \\[2pt]

\midrule

\multicolumn{5}{l}{\textbf{Global parameters for FGK rotation} $\theta_{\rm rot}$} \\[4pt]
\midrule

$p_{\rm rot}$
& $\mathrm{B}(2,5)$ 
& $0.20$ 
& $0.16$ 
& -- \\[2pt]

$\Delta A_{\rm Li, rot}$
& $\Gamma(\alpha=20,\;\beta=10)$
& $1.90$
& $0.45$
& dex \\[2pt]

$\sigma_{A_{\rm Li},{\rm rot}}$ 
& $\Gamma(\alpha=2,\;\beta=4)$ 
& $0.25$ 
& $0.35$ 
& dex \\[2pt]

$\alpha_{\rm FGK}$
& $\Gamma(\alpha=2,\;\beta=0.5)$ 
& $2.00$ 
& $2.82$
& -- \\

\midrule
\bottomrule
\end{tabular}
\label{tab:chronos_priors}
\tablefoot{
The priors are organised into four groups: (i) true stellar effective temperatures (local parameters, one for each of the $N$ objects), (ii) main global parameters for age and intrinsic $A_{\rm Li}$ dispersion, (iii) global parameters describing the transition between FGK and UCD/M regimes, and (iv) global parameters governing $A_{\rm Li}$ enhancement associated with stellar rotation. For each prior distribution we report its analytical mode and standard deviation ($\sigma$). Definitions are provided in Appendix \ref{ap:distributions}.}
\end{table}

\subsection{Multilayer perceptron}\label{sc:mlp}

\noindent We used a multilayer perceptron (\texttt{MLPs} code) to predict the theoretical lithium abundance, $A_{\rm Li}$, as a function of stellar age and effective temperature. Within the \textsc{Chronos} model, the NN provides the mean theoretical lithium abundance $\mu_{A_{\rm Li},i}$, which enters the latent lithium distribution defined in Eq. (\ref{eq:latent_two_component_mixture}).

The NN was trained on a grid of 456 BT-Settl evolutionary models \citep{Allard2012}, spanning ages from $5$ to $600$ Myr, $T_{\rm eff}$ from $\sim1200$ to $\sim7000$ K, and stellar masses from 0.01 to $1.4\,M_\odot$. The NN was implemented as a fully connected MLP with two input neurons (Age and $T_{\rm eff}$), four hidden layers with 64 neurons each and \texttt{ReLU} activations, and one output neuron with a sigmoid activation to predict $A_{\rm Li}$. Details of the preprocessing, training, and validation are provided in Appendix \ref{ap:mlp}. From the NN analysis, we remark that the typical prediction error, estimated by comparison with the BT-Settl grid, is $\sim$0.03--$0.05$ in scaled units, corresponding to $\sim$0.15--$0.25$ dex in $A_{\rm Li}$.

The BT-Settl evolutionary models adopted in this work were computed at solar metallicity. While the present validation was performed on a near-solar metallicity cluster \citep[e.g.][]{Soderblom2009}, the LDB technique is known to be only weakly sensitive to modest metallicity variations over the range of nearby young clusters. Theoretical studies indicate that modest changes in input physics, including metallicity variations typical of nearby young clusters, produce only minor shifts in LDB ages \citep{Burke2004}. Furthermore, empirical analyses of Li depletion in clusters with $-0.29\lesssim\mathrm{[Fe/H]}\lesssim0.18$ find no strong systematic dependence of Li depletion on metallicity \citep{Jeffries2023}. Within this regime, metallicity effects are sub-dominant compared to the primary age dependence of lithium depletion, justifying the use of solar-metallicity models at the level of precision aimed for.

\section{Data}\label{sc:data}

\noindent The applicability of Bayesian hierarchical models to stellar age inference improves in line with the availability of high-quality observational data. Due to its proximity, youth, and extensive observational history, the \object{Pleiades} cluster constitutes one of the best-studied benchmarks. Pleiades is a nearby ($\sim$136 pc) open cluster with a rich observational legacy and is widely used for lithium-based age determinations \citep{Bouy2015}. Age estimates based on the LDB typically converge around $\sim$125 Myr \citep[e.g.][]{Rebolo1992,Magazzu1993,Soderblom1993}, while evolutionary models, including rotation, can yield ages spanning $\sim$110--160 Myr depending on the adopted physics \citep{Barrado2016}. These considerations motivate the construction of a temperature-homogeneous dataset across the LDB regime ($\lesssim 3700$ K), where age sensitivity is strongest.

\subsection{Compilation of lithium and effective temperatures}

\noindent To maximise completeness around the LDB and ensure a uniform effective temperature scale derived using a consistent SED-fitting methodology (VOSA) and the same grid of BT-Settl atmospheric models, we compiled a master list of lithium detections by combining three sources: (i) the Pleiades lithium sample of $148$ sources from \citet[hereafter B+18]{Bouvier2018}, (ii) the DANCe.II lithium sample from \citet[hereafter D+16]{Barrado2016} with lithium measurements and effective temperatures derived with VOSA \citep{Bayo2008} from DANCe.I membership \citep{Bouy2015}, and (iii) a VOSA-based compilation of 213 Pleiades members with lithium detections and homogeneous $T_{\rm eff}$ estimates from our group. We note that PELS 069, from B+18, lacks a counterpart in the DANCe.I catalogue and was excluded from the dataset.

A key practical issue is that the coolest objects defining the lower-$T_{\rm eff}$ side of the LDB are predominantly characterised by VOSA-based $T_{\rm eff}$ \citep{Bayo2008,Barrado2016}, whereas a subset of warmer stars in B+18 adopts colour--$T_{\rm eff}$ relations based on \citet{Pecaut2013}. To avoid a heterogeneous $T_{\rm eff}$ scale precisely where the LDB constraint is most informative, we adopted VOSA-derived $T_{\rm eff}$ whenever available. For objects lacking VOSA solutions, we derived new VOSA-based estimates using the same configuration. This included, after quality and membership cuts (see Sect. \ref{sc:data:memb}), 10 sources from B+18 and 4 sources from D+16, not included in the initial VOSA sample of 213 objects. For the 10 B+18 sources, $T_{\rm eff}$ is computed using available multi-band photometry from Gaia DR3 $G$, $G_{\rm BP}$, and $G_{\rm RP}$ \citep{GaiaDR3}, 2MASS $JHK_{\rm s}$ \citep{2MASS}, four WISE bands \citep{WISE}, and, when available, Pan-STARRS $gri$ bands \citep{PanSTARRS}. We note that only three sources (with $T_{\rm eff} \gtrsim 4500\,\mathrm{K}$) lack Pan-STARRS $g$-band; two of them are detected only in $i$ (HII1593, V1289 Tau) and one only in $r$ (DH146), even though they have sufficient near-infrared and mid-infrared coverage for robust SED fitting. The four additional D+16 sources (BPL163, BPL327, LZJ50, PPL14), with $T_{\rm eff} \gtrsim 2700\,\mathrm{K}$, have all the photometric bands available and have also been reprocessed within the same VOSA framework to ensure full consistency in the $T_{\rm eff}$ scale across the dataset.

We also note that one object in common between catalogues (HCG332 / V348 Tau) exhibits discrepant temperature estimates across the literature and atypical behaviour with respect to distance and lithium. Therefore, it was treated as an outlier and removed from the sample.

\subsection{Membership and quality cuts}\label{sc:data:memb}

\noindent We applied a sequence of quality cuts to minimise contamination and remove unresolved multiples that can bias both photometry-based $T_{\rm eff}$ and $A_{\rm Li}$. First, sources without a reported Li\,\textsc{I} 6708 \AA{} equivalent width (EW) were removed. Second, known multiple systems were excluded using SIMBAD classifications, complemented with Gaia DR3 multiplicity indicators following \citet{Cifuentes2025}.

In addition, we rejected sources with evidence of close neighbours within $\sim$4\arcsec\ that could contaminate the photometry. This mitigates the impact of crowding and flux contamination on the SED fitting. Finally, we required a DANCe membership probability of $p_{\rm mem}>0.5$ \citep{Sarro2014,Bouy2015}. After applying these filters, we obtained a sample with $122$ sources.

\subsection{Lithium abundances and uncertainties}

\noindent For $T_{\rm eff}\gtrsim3000$ K, we can derive $A_{\rm Li}$ from Li\,\textsc{I} 6708\,\AA\ EWs using the curves of growth (CoG) from \citet{Franciosini2022}, interpolating in $(T_{\rm eff},\log g,{\rm EW}_{\rm Li})$ and adopting solar metallicity. Surface gravities follow \citet{Pecaut2013}. To ensure internal consistency across the FGK regime, $A_{\rm Li}$ uncertainties are propagated numerically through the local derivative $\mathrm{d}A_{\rm Li}/\mathrm{d}{\rm EW}_{\rm Li}$, and $T_{\rm eff}$ uncertainties from VOSA SED fitting are $\sim50$ K. We recompute $A_{\rm Li}$ for the B+18 objects using the \citet{Franciosini2022} CoG. For objects from D+16 with $T_{\rm eff}\lesssim3000$ K, we adopt the published $A_{\rm Li}$ values. For the remaining UCDs from B+18, we derive $A_{\rm Li}$ following the same prescription adopted in D+16 for the UCD regime, ensuring methodological consistency. In both catalogues, EW upper limits are reported for 22 sources. These are converted into upper limits in $A_{\rm Li}$ using the same CoG and treated as censored values within the bounded range $[-2.0,\,3.2]$ dex. Most of these objects (18 out of 22) lie at $T_{\rm eff}\lesssim3700$ K, with 14 sources located along the steep LDB, where the $A_{\rm Li}$–$T_{\rm eff}$ relation is nearly vertical and the models already predict strong depletion. In this regime, values below the detection limit are effectively indistinguishable, since the theoretical curves saturate at the depleted branch. As a result, the exact value below the upper limit has negligible impact on the inferred age. The remaining sources are in the Li-rich plateau ($T_{\rm eff}\lesssim2800$ K; 4 sources), and around $T_{\rm eff}\sim4000$ K and $A_{\rm Li}\sim0$ dex (4 sources). For sources with EW lacking reported uncertainties (including upper limits and 6 non-upper limit sources) we adopt a representative EW error equal to the mean uncertainty of the observed sample to enable consistent propagation into $A_{\rm Li}$. The final compiled sample provides, for each source, homogeneous $T_{\rm eff}$ and $A_{\rm Li}$ to minimise heterogeneity across the LDB, associated uncertainties, and ancillary astrometry.

\section{Model checking and results}\label{sc:validation}

\noindent In this section, we assess the statistical performance and consistency of the \textsc{Chronos} BHM. We first evaluate its behaviour on controlled synthetic datasets to test the internal consistency. We then validate the model on actual observations of the Pleiades cluster, examining chain convergence, residuals, and predictive and posterior performance.

\subsection{Sensitivity analysis on synthetic stellar associations}

\noindent To measure the accuracy and prior sensitivity of the age inference in \textsc{Chronos}, we used three controlled synthetic stellar associations generated from the same hierarchical model (Appendix \ref{ap:fake_data}). For each synthetic dataset, we fix $\mathrm{Age}_{\rm input} = 20$, $120$, and $400$ Myr, with a controlled $\sigma_{\mathrm{Age}}$ (fixed at 5\% of $\mathrm{Age}_{\rm input}$). The datasets were then analysed using the priors listed in Table \ref{tab:chronos_priors} to verify that the inference recovers the input age (Appendix \ref{ap:fake_data_results}).

\begin{figure}[!t]
  \centering

  \makebox[0.945\columnwidth][r]{%
    \includegraphics[width=0.9295\columnwidth,
      trim=0 1.655cm 0cm 1.5cm, clip]{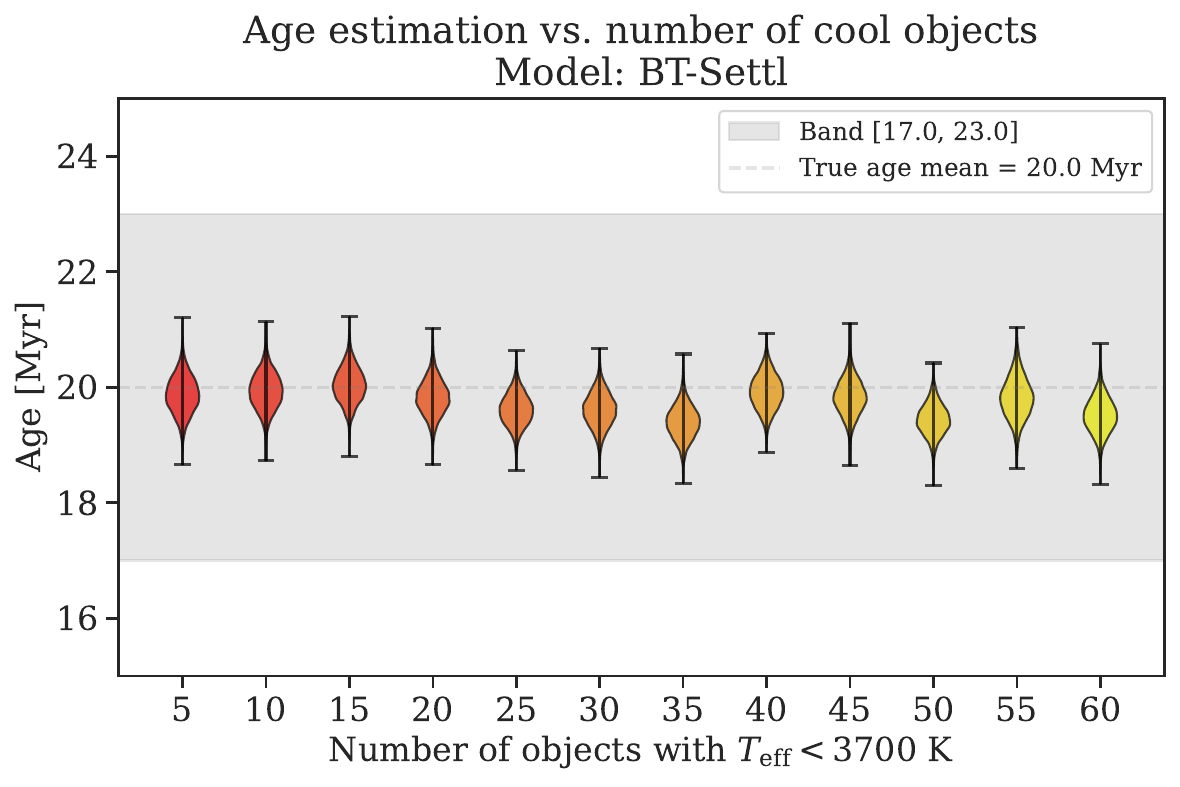}%
  }

  \vspace{1pt}

  \makebox[0.945\columnwidth][r]{%
    \includegraphics[width=0.945\columnwidth,
      trim=0 1.655cm 0cm 0.25cm, clip]{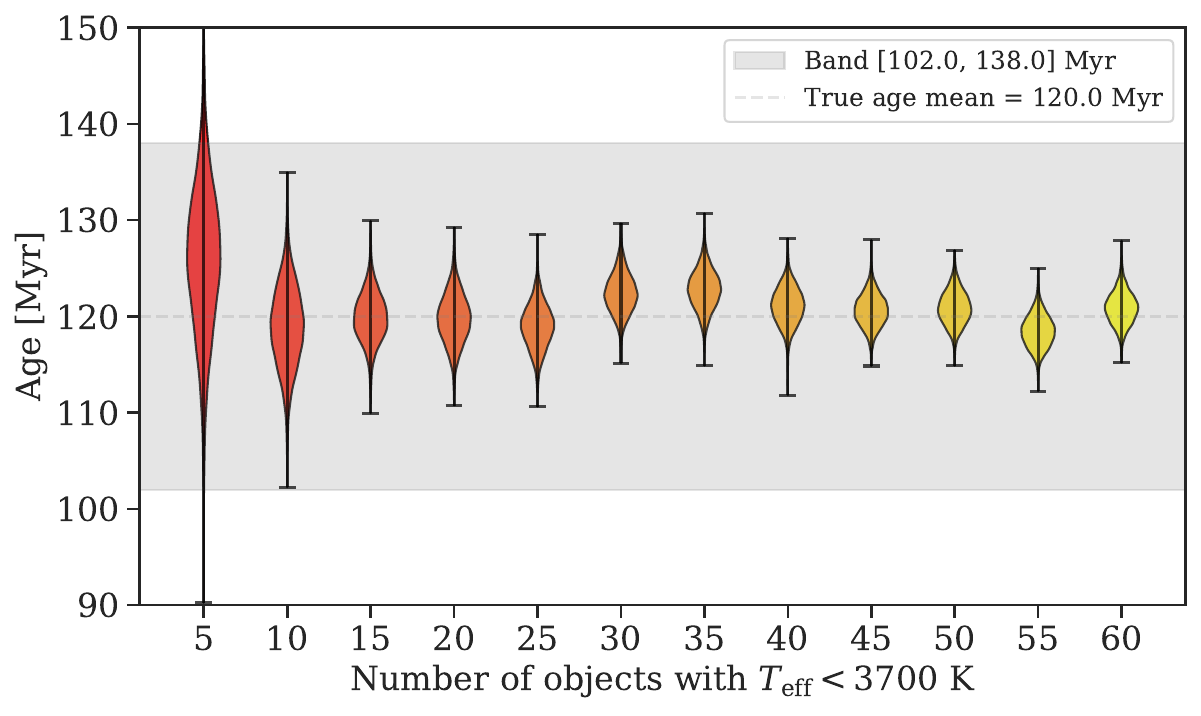}%
  }

  \vspace{1pt}

  \makebox[0.945\columnwidth][r]{%
    \includegraphics[width=0.945\columnwidth,
      trim=0 0cm 0cm 0.25cm, clip]{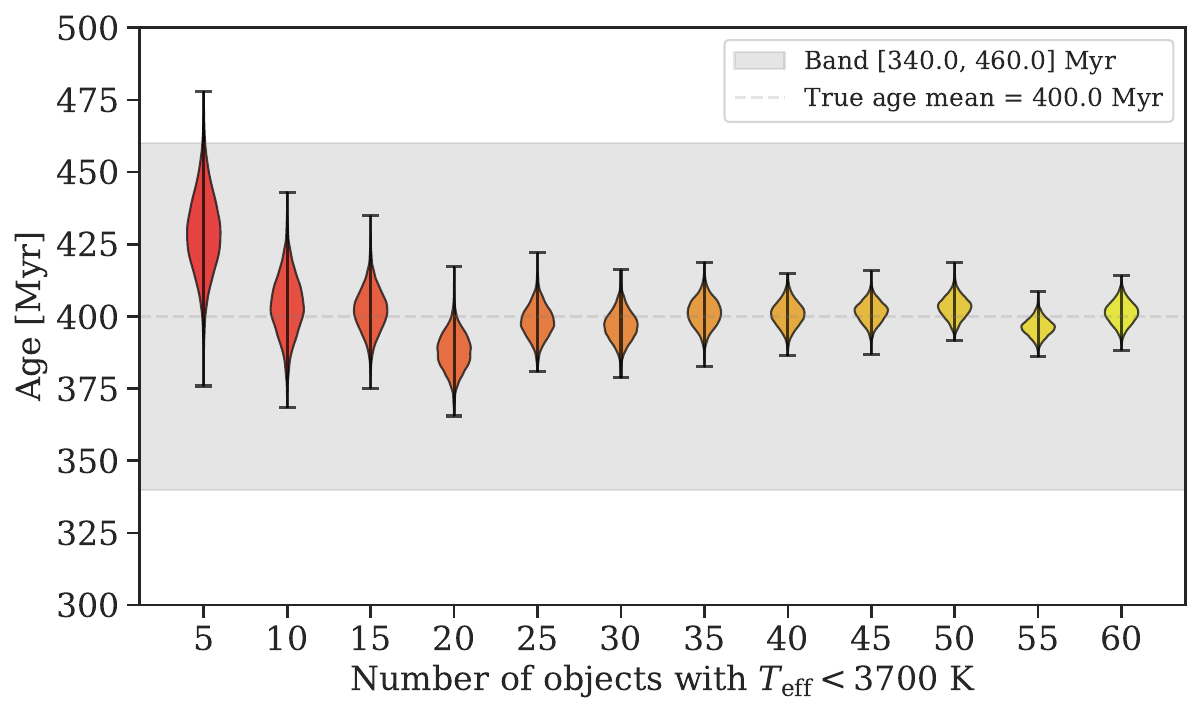}%
  }

  \caption{Violin plots of the posterior LDB age for 20, 120 and 400 Myr synthetic stellar associations. The total number of sources is fixed to 122 while we increase the number $n$ of objects at $T_{\rm eff}<3700$\,K. Whiskers indicate the central 99.7\% credible interval (HDI), and the dashed marker shows the posterior median.}
  \label{fig:general}
\end{figure}

To quantify the ability of the model to recover the cluster age and to test its internal consistency, we constructed 12 realisations with a fixed total sample size of 122 sources (matching the Pleiades dataset size) while varying the number of objects populating the LDB-sensitive regime ($T_{\rm eff}<3700$ K). Even with only 5--10 objects in this range, the inferred cluster age remains consistent with $\mathrm{Age}_{\rm input}$, with the posterior distributions lying within the $\mathrm{Age}_{\rm input}\pm 3\sigma_{\mathrm{Age}}$ interval in all realisations (shaded band in Fig. \ref{fig:general}). We further verified that the input mean age is recovered, all MCMC chains converged satisfactorily and the posterior predictive distributions are consistent with the simulated data.

An important outcome of this experiment is that the inferred age uncertainties are comparable to those obtained with lithium-based methods applied to clusters of similar age. In particular, for the Pleiades (Sect. \ref{sc:posterior}), the 68\% credible interval we obtain ($\sim 2.5\%$) is consistent with published LDB age uncertainties.

\subsection{Validation on the Pleiades dataset}

\noindent We applied \textsc{Chronos} to the Pleiades cluster to address three aspects: (i) numerical convergence of the MCMC posterior sampling; (ii) posterior structure and parameter identifiability; and (iii) predictive adequacy of the fitted generative model.

We first examined the standard MCMC diagnostics to check the convergence and sampling efficiency, and then analyse the posterior distributions of the global parameters to evaluate parameter constraints and potential dependencies. We also assessed the adequacy of the Gaussian measurement model with residual diagnostics. Predictive calibration is evaluated using probability integral transform (PIT) diagnostics under $K$-fold cross-validation and leave-one-out influence measures \citep{Vehtari2016}.

Finally, posterior predictive checks (PPCs; \citealt{Guttman1967,Rubin1984}) were used to assess whether replicated datasets drawn from the posterior distribution reproduce the global structure of the observed data.

\subsubsection{Convergence and stability}

\noindent Posterior inference was performed with MCMC using the No-U-Turn Sampler \citep[NUTS; ][]{hoffman2014nuts} as implemented in \texttt{PyMC} \citep{abrilpla2023pymc}. We ran $N_{\rm chains}=4$ independent chains with $N_{\rm tune}=10\,000$ tuning steps and $N_{\rm draw}=5\,000$ retained posterior draws per chain, resulting in $60\,000$ posterior samples in total\footnote{For the Pleiades (122 objects), the multi-chain parallel run requires $\sim$35 min on an Intel(R) i7 @ 3.60 GHz, 4 cores, and 32 GB of RAM.}. We used the default \texttt{PyMC} settings, and we increased the \texttt{target\_accept} NUTS parameter to 0.90--0.97 to minimise divergences and improve the convergence of the chains. We report posterior summaries and standard MCMC convergence diagnostics (e.g. $\hat{R}$, ESS, MCSE, and HDIs), computed with \texttt{ArviZ} \citep{arviz2019} under the prior configuration described in Table \ref{tab:chronos_priors}. Table \ref{tab:convergence_global} summarises convergence diagnostics for the global parameters, including the Gelman--Rubin statistic, effective sample size (ESS) in the bulk, posterior modes, and Monte Carlo standard errors (MCSE). All parameters exhibit $\hat{R}\simeq 1$ and large ESS, indicating good mixing and convergence. In particular, the inferred age is well constrained and numerically stable, providing a solid basis for the subsequent analysis. This confirms that the posterior results discussed below are not dominated by numerical sampling issues.

\subsubsection{Posterior results}\label{sc:posterior}

\begin{figure}[!ht]
\centering
\includegraphics[width=0.875\columnwidth]{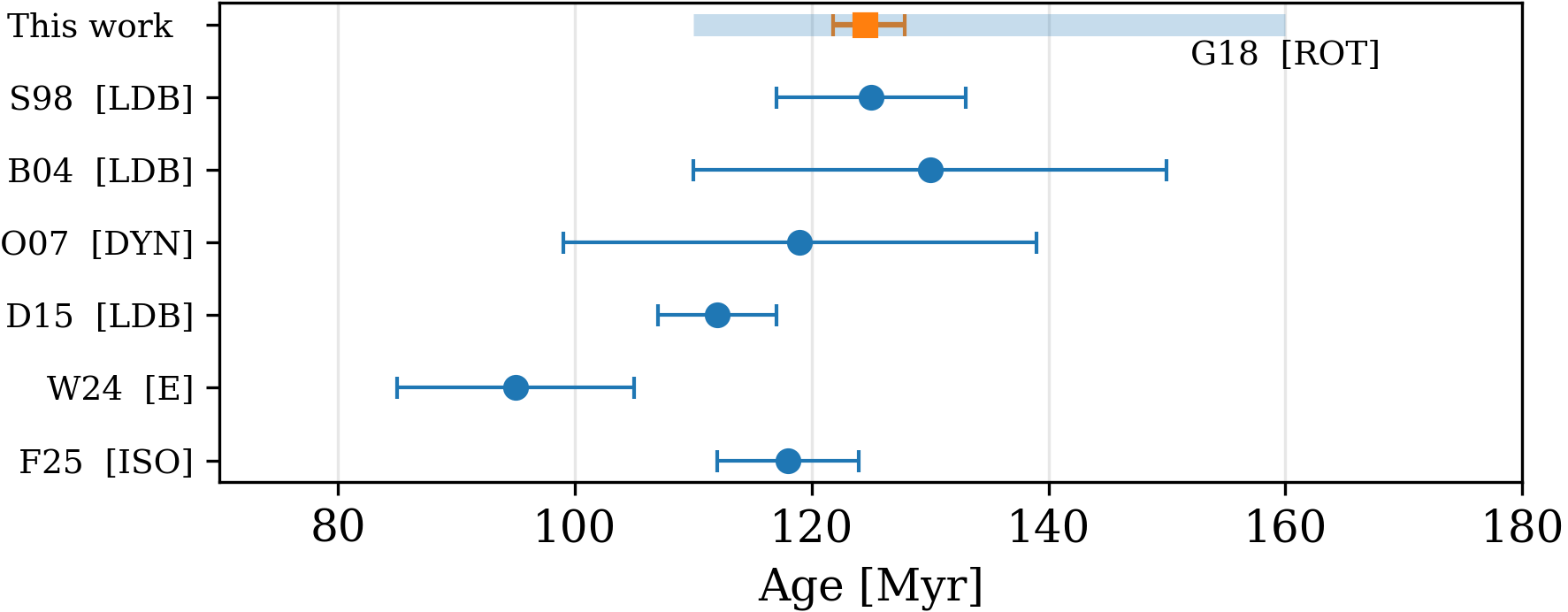}
\caption{Published Pleiades ages compared with \textsc{Chronos}. Error bars indicate uncertainties. \textbf{Method codes:} ISO = Isochrones fitting with \textsc{eagles}. LDB = Classical LDB. 
DYN = Dynamical. E = \textsc{eagles}. ROT = rotating stellar models \citep{Gossage2018}. \textbf{References:} Initial and last two digits of the publication year \citep{Stauffer1998,Barrado2004,Ortega2007,Dahm2015,Gossage2018,Weaver2024,Frasca2025}.}
\label{fig:pleiades_age_comparison}
\end{figure}

\noindent The LDB age inferred for the Pleiades with \textsc{Chronos} is consistent with the range of ages reported in the literature from independent methods (see Fig. \ref{fig:pleiades_age_comparison}). Notably, \citet{Frasca2025} applied the \textsc{eagles} model to fit empirical lithium depletion isochrones directly to $T_{\rm eff}$ and EW measurements over 3000--6500 K, obtaining an age of $\sim118\pm6$ Myr (or $\sim122\pm6$ Myr including upper limits). While the central values are in good agreement, the posterior uncertainty obtained with \textsc{Chronos} is smaller than in recent LDB-based estimates. This reflects differences in our methodology, where the age is constrained by the full $(T_{\rm eff}, A_{\rm Li})$ distribution and the probabilistic modelling of intrinsic dispersion and rotation effects.

\FloatBarrier

\begin{figure*}[!ht]

\begin{tikzpicture}
  \node[anchor=south west, inner sep=0] (img) at (0,0)
    {\includegraphics[width=\textwidth,trim=0 0.35cm 0 0.25cm,clip]{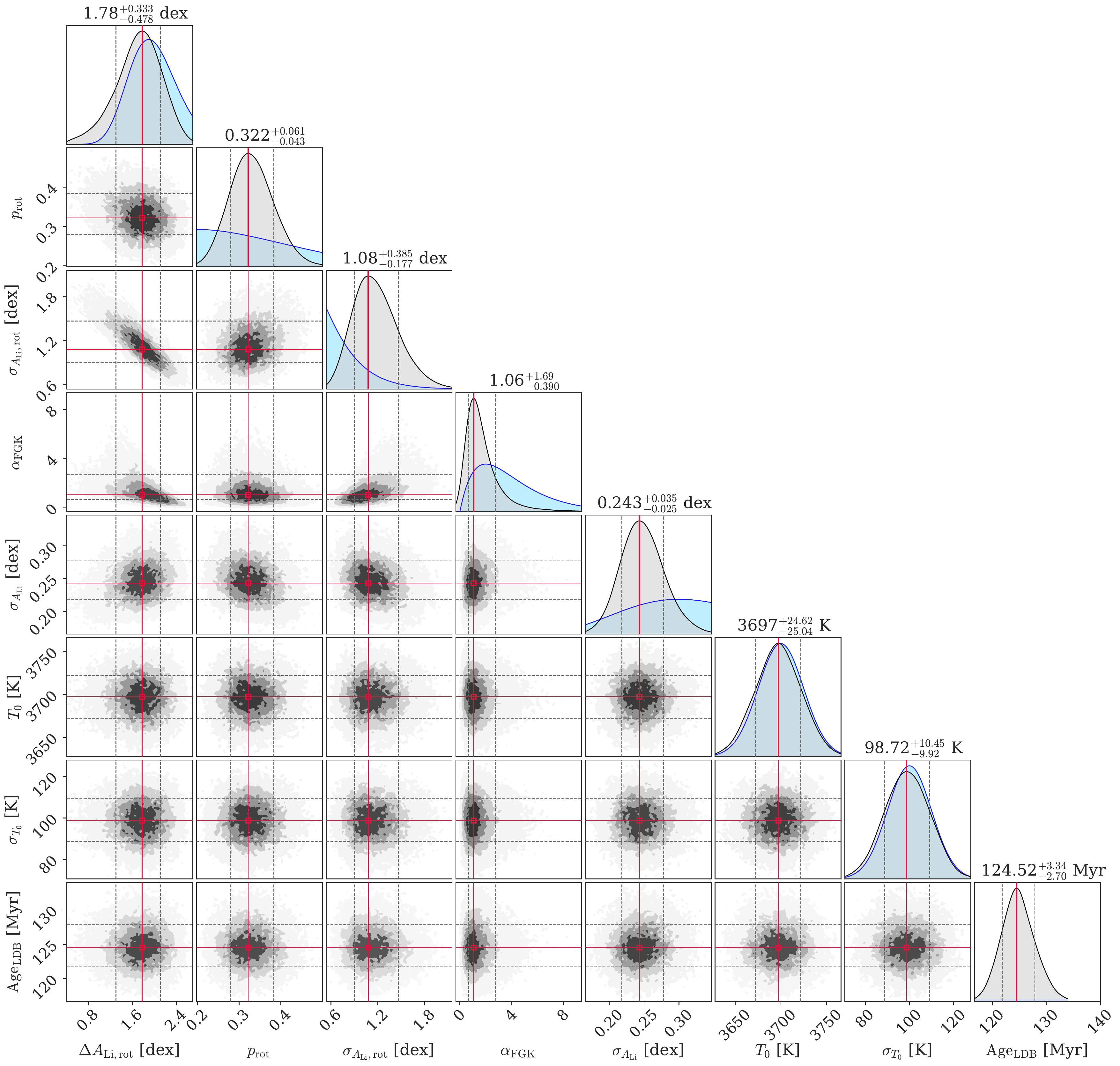}};
\end{tikzpicture}
\caption{Corner plot of the posterior distributions for each global parameter from $\Theta$. Red lines indicate the posterior mode and asymmetric uncertainties from the 95\% HDI, computed with \texttt{ArviZ} \citep{arviz2019}. The diagonal panels also display the prior distributions (blue) overlaid with the posterior densities (grey), enabling a direct visual comparison with prior assumptions.}
\label{fig:corner_global}
\end{figure*}

The modes and $\sigma$ of the prior distributions (Table \ref{tab:chronos_priors}) enable a quantitative comparison with the modes in Table \ref{tab:convergence_global} and clearly illustrated in the diagonal panels from Fig. \ref{fig:corner_global}. Based on this, the posterior of the age and rotation parameters (Fig. \ref{fig:corner_global}) exhibits substantially smaller dispersion than their respective priors, indicating that these parameters are informed by the data rather than prior assumptions. In contrast, the transition parameters remain regularised by their physically motivated priors (Appendix \ref{ap:w_spec}). The joint posterior (see Fig. \ref{fig:corner_global}) reveals a moderate correlation between $\Delta A_{\rm Li, rot}$ and $\sigma_{A_{\rm Li}, \rm rot}$ (first column and second row of Fig. \ref{fig:corner_global}). This behaviour reflects the fact that both parameters contribute to modelling lithium enhancement and describe the same physical effect rather than a structural degeneracy of the model. 

In the adopted mixture from Eq. \ref{eq:latent_two_component_mixture}, both parameters, $\Delta A_{\rm Li,rot}$ and $\sigma_{A_{\rm Li},\rm rot}$, contribute to the effective width and offset of the fast rotators distribution. Additional tests in which the dispersion term $\sigma_{A_{\rm Li,rot}}$ is suppressed show that the model naturally compensates by increasing and broadening $\Delta A_{\rm Li,rot}$, which absorbs the variance previously captured by $\sigma_{A_{\rm Li,rot}}$. This demonstrates that the two parameters control a common effective degree of freedom, and that their coupling is therefore expected. Importantly, this redundancy does not affect the inference of the LDB age. The posterior distribution of $\mathrm{Age}_{\rm LDB}$ remains essentially uncorrelated with the rotation parameters, indicating that the age determination is driven by the morphology of the LDB rather than by the parametrisation of rotation-induced lithium enhancement.

\subsubsection{Residuals on model predictions}\label{sc:residuals}

\noindent To assess the adequacy of the Gaussian measurement model, we analyse standardised residuals for $A_{\rm Li}$ and $T_{\rm eff}$. We define, for each source, $i$, observable, $j\in\{A_{\rm Li},T_{\rm eff}\}$, and posterior draw, $s$ via
\begin{equation}
r_{i,j}^{(s)} =
(y_{i,j}^{\rm obs} - y_{i,j}^{(s)})/\sigma_{i,j,{\rm obs}},
\end{equation}
where $\sigma_{i,j,{\rm obs}}$ is the reported observational uncertainty, and $y_{i,j}^{(s)}$ is the posterior sample entering the likelihood at draw $s$.

\begin{figure}[!ht]
\centering
    \makebox[0.925\columnwidth][c]{%
        \hspace*{-0.35cm}%
        \begin{overpic}[width=0.409\textwidth,trim=0 0.25cm 0 0.25cm,clip]{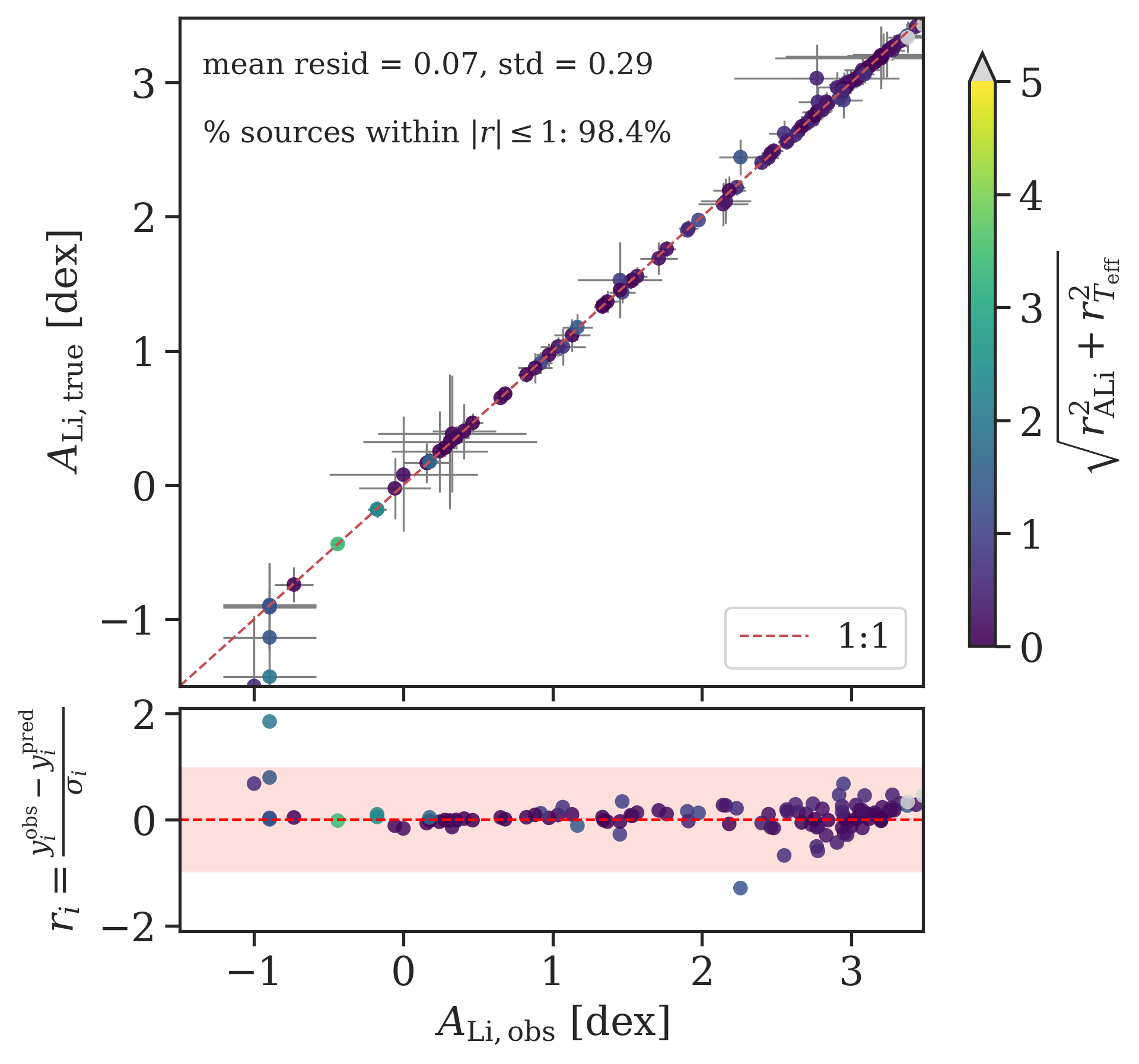}
            \put(17.5,71.25){\textbf{(a)}}
        \end{overpic}
    }

    \makebox[0.925\columnwidth][c]{%
        \hspace*{-0.35cm}%
        \begin{overpic}[width=0.4125\textwidth,trim=0.15cm 0.25cm 0 0.25cm,clip]{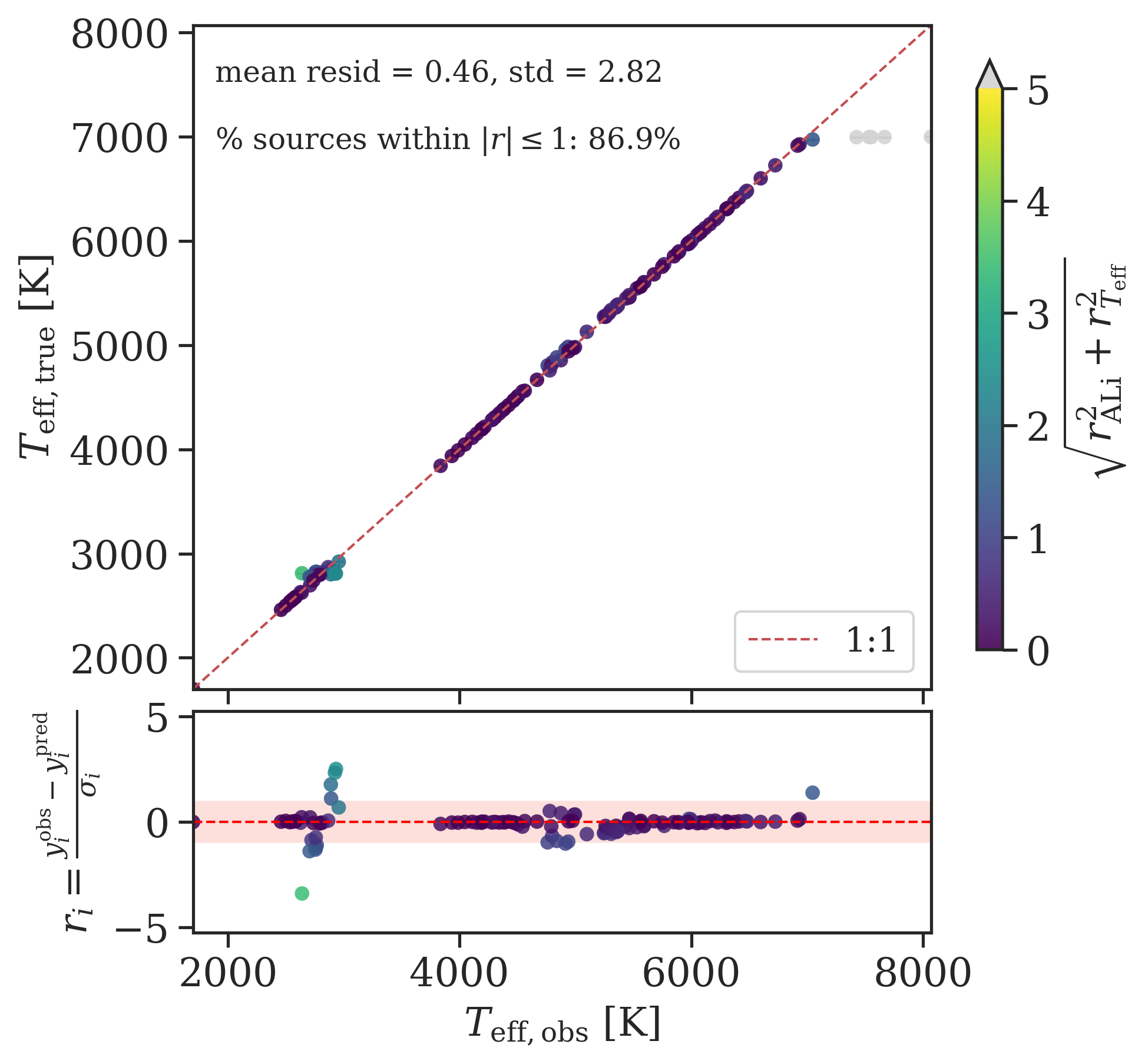}
            \put(18.5,70.25){\textbf{(b)}}
        \end{overpic}
    }

    \caption{Residuals for $A_{\rm Li}$ \textbf{(a)} and $T_{\rm eff}$ \textbf{(b)}. For each figure, top: latent posterior mean versus observed value, with the 1:1 relation, and bottom: per-source residuals and $|r|\leq 1$ band. Five stars ($T_{\rm eff}\gtrsim 7000\,{\rm K}$), with $|r|>5$, are coloured grey, and lie outside the NN $T_{\rm eff}$ range; their residuals do not reflect model deficiencies. We omit these sources from the bottom panel for clarity.}
    \label{fig:res}
\end{figure}

The upper panels of Fig. \ref{fig:res} display the latent posterior mean versus the observed values, while the lower panels show per-source residuals, both for $A_{\rm Li}$ and $T_{\rm eff}$. In the upper panels, we also display error bars to visualise the relative contribution of measurement and posterior uncertainty in the latent variables. Horizontal error bars correspond to $\sigma_{i,j,{\rm obs}}$, while vertical error bars represent the posterior uncertainty of the latent variable, computed as the 68\% HDI for each per-source latent distribution. In the case of $T_{\rm eff}$, both observed and posterior error bars are smaller than the marker size.

\begin{figure}[!t]
  \centering

  \begin{minipage}{0.4275\columnwidth}
    \begin{overpic}[width=\linewidth,trim=0cm 0.25cm 0 0.125cm,clip]{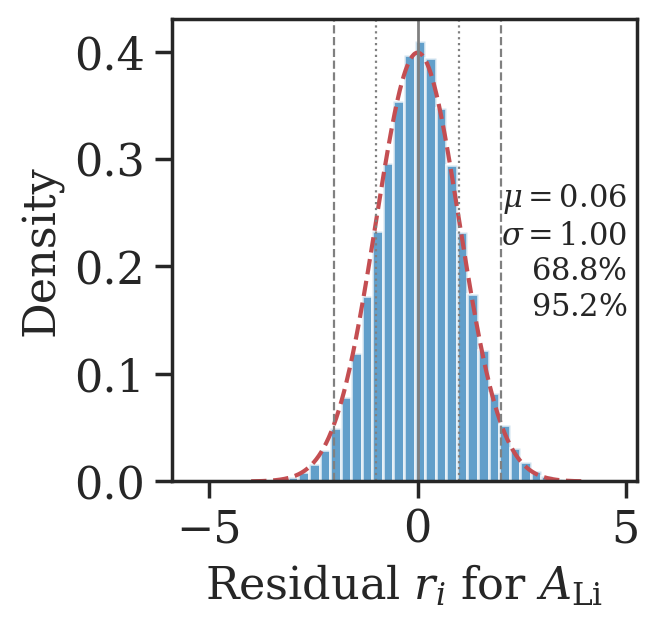}
      \put(30,80){\textbf{(a)}}
    \end{overpic}
  \end{minipage}%
  \hspace{0.04\columnwidth}%
  \begin{minipage}{0.385\columnwidth}
    \begin{overpic}[width=\linewidth,trim=0.9cm 0.25cm 0 0.125cm,clip]{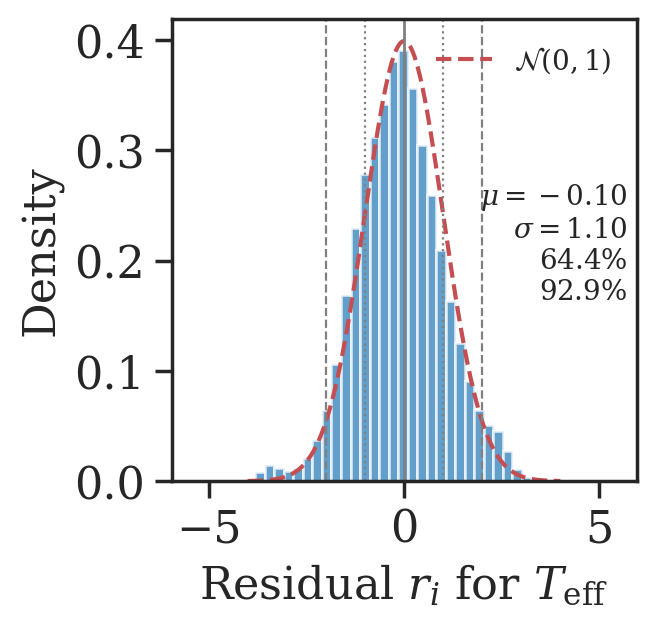}
      \put(21.5,83.5){\textbf{(b)}}
    \end{overpic}
  \end{minipage}

  \caption{Distribution of draw-wise standardised residuals for $A_{\rm Li}$ \textbf{(a)} and $T_{\rm eff}$ \textbf{(b)}, pooled over sources and posterior draws. The dashed curves show the reference $\mathcal{N}(0,1)$ distribution.}
  \label{fig:hist}
\end{figure}

The associated histograms (Fig. \ref{fig:hist}) show the pooled draw-wise residuals, combining all sources and posterior draws, and compare them with the reference normal distribution $\mathcal{N}(0,1)$. For $A_{\rm Li}$, the residual distribution is consistent with $\mathcal{N}(0,1)$ ($\mu \simeq 0.06$, $\sigma \simeq 1.00$). For $T_{\rm eff}$, the dispersion is larger ($\sigma \simeq 3$) when all objects are included, driven by 5 objects with extreme residuals ($|r|\gtrsim5$). These correspond to the hottest stars in the sample ($T_{\rm eff}\gtrsim7000\,\mathrm{K}$), which lie outside the $T_{\rm eff}$ range covered by the NN. In this regime, the predicted $T_{\rm eff}$ saturates at the boundary of the training range, leading to artificially large residuals (see Appendix \ref{ap:residuals_z}). To assess their impact, we repeated the analysis excluding these 5 sources (\textbf{b} panel of Fig. \ref{fig:hist}). In this case, the $T_{\rm eff}$ residual distribution becomes consistent with $\mathcal{N}(0,1)$, with only a mild residual asymmetry due to a few sources with $r\sim -3$. We emphasise that these objects lie outside the $T_{\rm eff}$ regime that constrains the LDB and therefore have negligible influence on the inferred age, remaining unchanged within uncertainties at $\mathrm{Age}\approx 125\pm2$. We conclude that these extreme residuals do not reflect a deficiency of the measurement model.

\subsubsection{Cross-validation}\label{sc:CV}

\noindent While residuals probe local measurement consistency, global predictive calibration is assessed using PIT-based diagnostics under cross-validation. Leave-one-out cross-validation (LOO-CV) based on Pareto-smoothed importance sampling (PSIS-LOO; \citealt{Vehtari2015}) provides an approximation to exact LOO-CV in many Bayesian models. However, PSIS can be unstable in hierarchical models where each observation is associated with well-constrained latent variables and mixture components \citep{Vehtari2016}. In our model, each star has latent variables for $T_{\rm eff}$ and $A_{\rm Li}$ tightly constrained by their own measurements. Removing one star thus alters not only the global posterior but also the latent variables associated with it. Such changes are difficult to approximate through PSIS. Therefore, we chose to compute LOO-reweighted posterior summaries for the global age, exclusively as an influence diagnostic, and define an age score via
\begin{equation}
z_{\rm age,i}
=
\left[\mu_{-i}(\mathrm{Age})-\mu(\mathrm{Age})\right]/\sigma(\mathrm{Age}),
\end{equation}
where $\mu(\mathrm{Age})$ and $\sigma(\mathrm{Age})$ are the posterior mean and standard deviation obtained from the full dataset and $\mu_{-i}(\mathrm{Age})$ is the LOO-reweighted posterior mean after down-weighting star $i$.

\begin{figure}[!ht]
    \includegraphics[page=2,width=0.975\columnwidth,trim=0 0.25cm 0 0.75cm,clip]{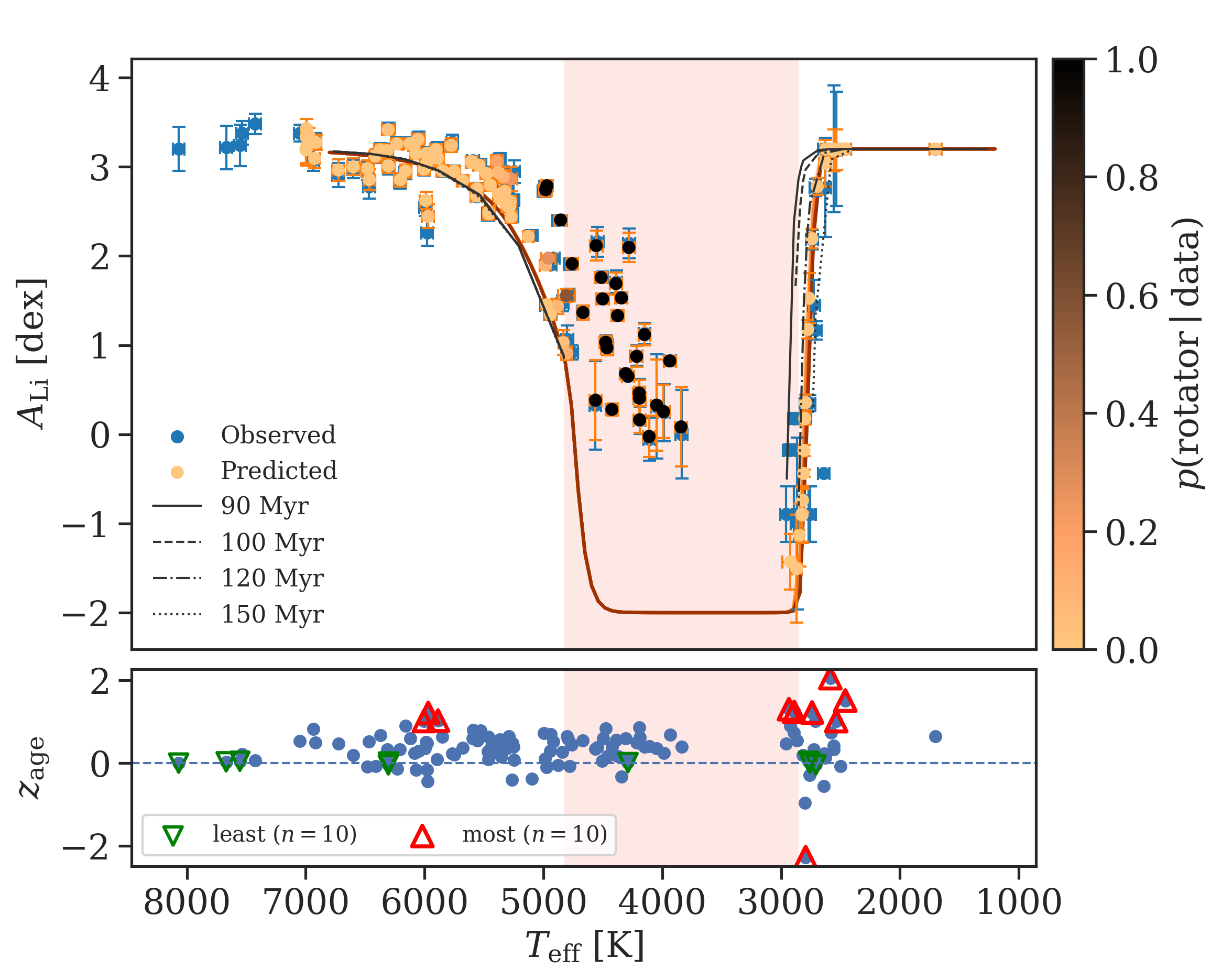}
    \caption{Lithium depletion diagram. Top: Observed (blue) and $A_{\rm Li,true}(T_{\rm eff,true})$ (coloured from orange to black, reflecting the posterior probability of the star being a fast rotator; see Appendix \ref{ap:prob_rot}). The red curve is the NN prediction of $A_{\rm Li} = f_{\rm NN}(\mathrm{Age},T_{\rm eff})$ for the inferred age. Black curves represent BT-Settl theoretical curves for ages around the inferred age. The flattening of the predicted curves in the fully depleted regime (shaded area) is an artefact of imposing a lower bound $A_{\mathrm{Li}}=-2$ dex on the $A_{\rm Li}$ NN predictions, despite the theoretical expectation $A_{\mathrm{Li}}\to -\infty$. Bottom: LOO age influence score $z_{\rm age}$ as a function of $T_{\rm eff}$. Triangles mark least and most influential sources.}
    \label{fig:LDD_z}
\end{figure}

The distribution of $z_{\rm age}$ is centred close to zero but shows a mild asymmetry (Fig. \ref{fig:LDD_z}), with $|z_{\rm age}|\lesssim2.5$ for all sources. The largest shifts are produced by objects located in the $T_{\rm eff}$ interval where the LDB exhibits the strongest age sensitivity. The slight preference for $z_{\rm age}>0$ arises from the asymmetric contribution of different regions of the $(T_{\rm eff},A_{\rm Li})$ plane to the age inference.

Overall, FGK stars tend to lie above the theoretical lithium curve due to rotation effects, which favour younger ages, particularly for sources outside the LDB (e.g. in the Li-rich plateau). In contrast, sources far from the LDB, such as those in the Li-rich plateau ($T_{\rm eff}\lesssim2700$ K), have very weak sensitivity to age. As a result, the age inference is effectively anchored by stars located near the steep LDB transition ($T_{\rm eff}\sim2800$ K), where small offsets in $A_{\rm Li}$ translate into significant age shifts. Removing individual stars that favour younger solutions therefore tends to shift the inferred age towards slightly older values, leading to $z_{\rm age}>0$.

\begin{figure}[!t]
  \centering
  \begin{minipage}{0.465\columnwidth}
    \begin{overpic}[width=\linewidth,page=1,trim=0cm 0.375cm 0 0.25cm,clip]{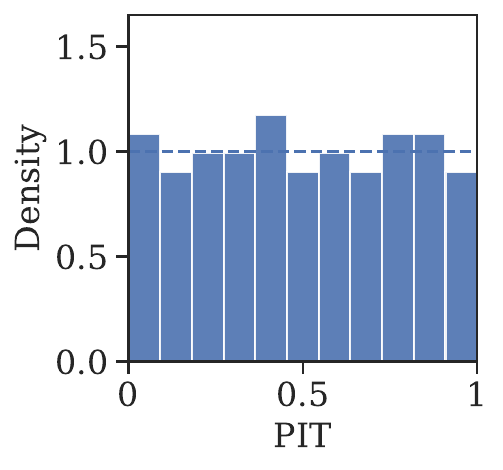}
      \put(30,75.25){\textbf{(a)}}
    \end{overpic}
  \end{minipage}%
  \hspace{0.04\columnwidth}%
  \begin{minipage}{0.465\columnwidth}
    \begin{overpic}[width=\linewidth,page=3,trim=0cm 0.375cm 0 0.25cm,clip]{loo_pit_KFOLD.pdf}
      \put(30,75.25){\textbf{(b)}}
    \end{overpic}
  \end{minipage}

  \caption{$K$-fold probability integral transform (PIT) histograms for the held-out observations under the refitted hierarchical model. Panel \textbf{(a)}: $T_{\rm eff}$. Panel \textbf{(b)}: $A_{\rm Li}$. The dashed line indicates $\mathcal{U}(0,1)$, as PIT is expected to be uniformly distributed on $[0,1]$ under the correct predictive calibration.}
  \label{fig:kfold_pit}
\end{figure}

\begin{figure}[!t]
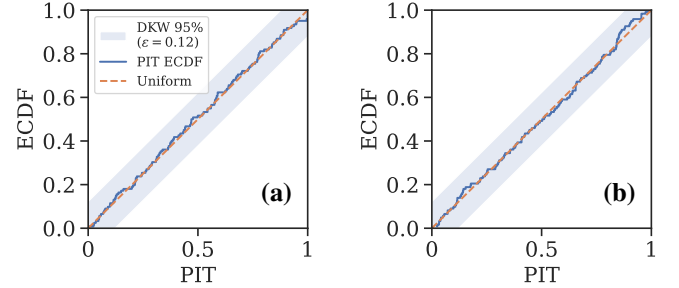

  \centering
  \begin{minipage}{0.465\columnwidth}
    \begin{overpic}[width=\linewidth,page=2,trim=0cm 0.375cm 0 0.25cm,clip]{loo_pit_KFOLD.pdf}
      \put(90,24.125){\makebox(0,0)[rb]{\textbf{(a)}}}
    \end{overpic}
  \end{minipage}%
  \hspace{0.04\columnwidth}%
  \begin{minipage}{0.465\columnwidth}
    \begin{overpic}[width=\linewidth,page=4,trim=0cm 0.375cm 0 0.25cm,clip]{loo_pit_KFOLD.pdf}
      \put(90,24.125){\makebox(0,0)[rb]{\textbf{(b)}}}
    \end{overpic}
  \end{minipage}

  \caption{Empirical cumulative distribution functions (ECDFs) of the $K$-fold PIT values for the held-out observations. Panel \textbf{(a)}: $T_{\rm eff}$. Panel \textbf{(b)}: $A_{\rm Li}$. The dashed diagonal line is the uniform CDF expected under perfect predictive calibration. The shaded region shows the DKW band.}
  \label{fig:kfold_pit_ecdf}
\end{figure}

For the predictive assessment, we adopted $K$-fold refitting, which provides a robust approximation to star-wise CV in hierarchical models \citep{Vehtari2016}. The dataset was partitioned into $K=10$ folds at the local level (see Appendix \ref{ap:PIT}). In this CV, each fold excluded all observables of the held-out sources and the model is refitted to the remaining data. Under correct predictive calibration, the corresponding PIT values can be expected to follow a uniform distribution in the $[0,1]$ interval. The resulting $K$-fold PIT histograms and empirical cumulative distribution functions (ECDFs) are consistent with uniformity within the 95\% Dvoretzky–Kiefer–Wolfowitz (DKW; Appendix \ref{ap:dkw}) bands (Figs. \ref{fig:kfold_pit} and \ref{fig:kfold_pit_ecdf}), indicating no detectable predictive miscalibration under star-wise CV. We further perform a PPC based on age using the LOO-PIT-reweighted test (see Appendix \ref{ap:loo_sanity}).

\subsubsection{Posterior predictive check}

\noindent We performed posterior predictive checks (PPCs; see Appendix \ref{ap:ppc}) to assess whether the model reproduces the empirical distributions of $T_{\rm eff}$ and $A_{\rm Li}$ under the posterior predictive distribution \citep[PPDs;][]{Guttman1967,Rubin1984,Gelman2013}.

\begin{figure}[!t]
  \begin{minipage}{0.5\columnwidth}
    \begin{overpic}[width=\linewidth,page=1,trim=0 0.385cm 0 0.275cm,clip]{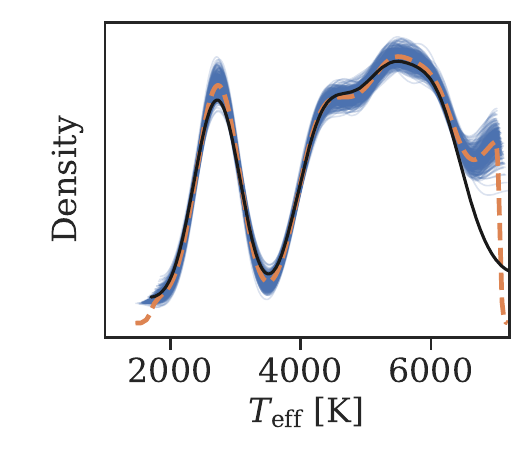}
      \put(92.5,25){\makebox(0,0)[rb]{\textbf{(a)}}}
    \end{overpic}
  \end{minipage}%
  \begin{minipage}{0.4016\columnwidth}
    \begin{overpic}[width=\linewidth,page=2,trim=1.75cm 0.385cm 0 0.275cm,clip]{ppc.pdf}
      \put(92.5,25){\makebox(0,0)[rb]{\textbf{(b)}}}
    \end{overpic}
  \end{minipage}

  \caption{PPCs for panel \textbf{(a)}: $T_{\rm eff}$ and panel \textbf{(b)}: $A_{\rm Li}$. Blue curves show PPD samples, the dashed orange line indicates the mean PPD, and the black curve corresponds to the observed distribution.}
  \label{fig:ppc_global}
\end{figure}

The PPCs are complementary to the CV (Sect. \ref{sc:CV}), where $K$-fold PIT evaluates out-of-sample calibration. Figure \ref{fig:ppc_global} shows PPCs for $A_{\rm Li}$ and $T_{\rm eff}$. The model reproduces the overall $A_{\rm Li}$ and $T_{\rm eff}$ distributions in the age-sensitive regime ($T_{\rm eff}\lesssim3700$ K). Deviations at the highest $T_{\rm eff}$ arise from stars near or beyond the NN training range ($\sim7000$ K). The predictive intervals are systematically wider than the data (e.g. the nominal 50\% interval contains $\sim85$--95\% of the observations; see Table \ref{tab:ppc_metrics}), indicating an over-dispersed PPD. We also note an overprediction at the Li-rich plateau ($A_{\rm Li}\sim3$ dex). This does not originate from the $A_{\rm Li,true}$ distribution, which is correctly reproduced, but from the observational level of the model. The Gaussian mapping from $A_{\rm Li,true}$ to $A_{\rm Li,obs}$ produces predictive realisations beyond the physical upper bound, leading to an accumulation of probability near the boundary in the PPC.

This effect on PPC for $A_{\rm Li,true}$ PPDs is further enhanced by marginalisation over the full population and the KDE used for visualisation, resulting in a broader lithium-rich peak. Since this behaviour is confined to a regime with weak age sensitivity, it does not affect the inferred LDB age and would, at most, lead to an overestimation of its uncertainty.

\section{Discussion}

\noindent The validation presented in Sect. \ref{sc:validation} shows that for the Pleiades dataset, the \textsc{Chronos} hierarchical model provides a statistically coherent description of the LDB and yields an LDB age estimate that is stable under multiple diagnostic tests. In this section we place these results in a broader context, discuss the main limitations of the current model, and compare our approach with recent lithium-based age-dating methods, in particular the \textsc{eagles} model \citep{Jeffries2023b,Weaver2024}.

\subsection{Model adequacy and inference robustness}

\noindent The joint posterior distributions indicate that the hierarchical model is well identified and appropriately regularised. In particular, the inferred LDB age is tightly constrained and exhibits only weak correlations with parameters describing rotation-induced $A_{\rm Li}$ dispersion. This indicates that the age is primarily driven by objects in the LDB $T_{\rm eff}$ regime, rather than by the prior. At the same time, no individual object dominates the inference.

The residual diagnostics and posterior predictive checks consistently support this conclusion. Lithium abundance residuals are centred near zero with no systematic trends, while the effective temperature residuals show a good agreement for the bulk of the sample. A negligible number of hot stars ($T_{\rm eff}\gtrsim7000$ K) exhibit larger residuals, as expected given the validity range of the neural network calibration, but they do not affect the temperature regime relevant for LDB age inference.

A key result of this work is provided by star-wise $K$-fold cross-validation, which indicates that the PPDs remain broadly consistent with calibration expectations under realistic out-of-sample conditions. This demonstrates that the inferred age is not driven by a small subset of influential sources, but reflects coherent information distributed across the stellar population. Taken together, the PPC results demonstrate that \textsc{Chronos} generates replicated datasets that are statistically consistent with the observed $T_{\rm eff}$ and $A_{\rm Li}$ distributions. This agreement supports the adequacy of the adopted likelihood and the explicit inclusion of rotation effects.

\subsection{Effect of rotation}

\noindent In \textsc{Chronos}, the parameter, $p_{\rm rot}$, is a phenomenological parameter describing the fraction of sources with $A_{\rm Li}$ enhancement, rather than a direct proxy for the rotational velocity distribution. Slowly rotating FGK stars whose rotation does not measurably affect surface lithium are therefore indistinguishable from non-rotators in this model.

The posterior correlation between $\Delta A_{\rm Li,rot}$ and $\sigma_{A_{\rm Li,rot}}$ suggests that rotation does not introduce a single deterministic lithium offset. Instead, this correlation gives rise to a distribution of enhancements across the FGK regime. This behaviour is consistent with recent observational results showing that lithium dispersion develops early in the PMS phase and increases in concert with rotation-dependent processes \citep{Jackson2025}. Accordingly, the rotation component in \textsc{Chronos} should be interpreted as an effective description of unresolved rotation-linked physics (such as magnetic activity, structural inflation, or surface inhomogeneities) rather than as a physical model of angular momentum evolution. The hierarchical formulation allows this diversity to be captured statistically without committing to a specific mechanism.

\subsection{Comparison with recent lithium-based age-dating models and \textsc{Chronos} BHM caveats}

\noindent Recent studies have highlighted that the physical origin of lithium dispersion relates to stellar rotation \citep[e.g.][]{Soderblom2014,Somers2015,Jeffries2021}. Although rotation is strongly correlated with lithium excess, reproducing both the amplitude and rotation dependence of the observed dispersion likely requires additional assumptions about magnetic activity and surface inhomogeneities that are not yet observationally well constrained, particularly during the PMS phase \citep{Jackson2025}. In this sense, the rotation in \textsc{Chronos} should not be viewed as a representation of any single physical mechanism.

\textsc{Chronos} shares this physical interpretation with recent empirical approaches such as \textsc{eagles} \citep{Jeffries2023b,Weaver2024}, but differs fundamentally in methodology. Whereas \textsc{eagles} relies on neural networks trained directly on observational lithium data to capture empirical correlations, \textsc{Chronos} adopts a generative hierarchical approach in which a neural network acts as a surrogate for theoretical lithium depletion curves, and empirical effects are incorporated probabilistically at the population level. These approaches are therefore complementary: \textsc{Chronos} is optimised for physically grounded population-level age inference in young clusters where the lithium depletion boundary provides strong constraints across a broad $T_{\rm eff}$ range (from UCDs to solar-type stars), while \textsc{eagles} offers empirical age estimates across a wider age range.

Despite the overall statistical performance demonstrated here, several caveats should be noted. The NN is trained on a specific set of theoretical models and inherits their systematic uncertainties, particularly near the edges of the calibrated $T_{\rm eff}$ range. Moreover, the present model assumes a single coeval population and treats rotation phenomenologically through its imprint on $A_{\rm Li}$, rather than via an explicit physical model of angular momentum evolution. As with any surrogate-based generative model, systematic uncertainties in the underlying calculations are not eliminated but propagated through the model levels.

\subsection{Future work}

\noindent The long-term goal of \textsc{Chronos} is to construct a self-consistent stellar age scale by combining multiple dating techniques within a unified Bayesian hierarchical model. Lithium-based ages provide a natural starting point for this effort, given their high precision and strong physical grounding at young ages, as demonstrated in this work. In the short term, we plan to apply \textsc{Chronos} to additional open clusters and moving groups spanning a broader range of ages and metallicities. This will enable a systematic assessment of model performance beyond the Pleiades and will test the robustness of the model under more diverse stellar populations and observational conditions.

We are currently working to include additional age-sensitive observables and complementary chronometers \citep{Olivares2025}, as well as metallicity as an additional dimension by training the neural network component on evolutionary models spanning a range of chemical compositions. A fully self-consistent hierarchical formulation would involve modelling lithium equivalent widths directly, introducing latent equivalent widths $\mathrm{EW}_{\rm true}$ and embedding the CoG transformation, $A_{\rm Li} = g(\mathrm{EW}, T_{\rm eff})$, within the probabilistic framework. This approach would allow a quantitative assessment of the influence of EW--$T_{\rm eff}$ covariance, when applied to stellar populations with non-solar abundances.

\section{Conclusions}

\noindent In this work, we present the first version of \textsc{Chronos}, a Bayesian hierarchical model designed to infer stellar population ages from the LDB in a statistically rigorous and physically interpretable way. Using the Pleiades cluster as a benchmark case, we have shown that the model is numerically stable, aptly identified, and robust under a comprehensive set of validation tests, including convergence diagnostics, residual analyses, PPCs, and star-wise $K$-fold CV. The inferred LDB age is tightly constrained and driven by physically informative regions of the data, rather than by prior assumptions. The explicit generative structure of \textsc{Chronos}, combining a NN trained on theoretical models with a probabilistic population-level treatment, allows empirical effects such as rotation-induced lithium dispersion to be incorporated in a controlled and interpretable manner. The resulting synthetic data generator reproduces the main morphological features of observed lithium patterns and provides a powerful tool for sensitivity analyses and methodological validation.

Overall, \textsc{Chronos} has been validated using a comprehensive suite of internal tests for lithium-based age inference, including parameter recovery, convergence diagnostics, residual analysis, posterior predictive checks, and star-wise cross-validation. The results demonstrate that lithium-based stellar chronology can be formulated as a coherent hierarchical inference problem, providing a statistically rigorous and physically interpretable framework. At present, \textsc{Chronos} constitutes a methodologically robust step towards the construction of a self-consistent and absolute stellar age scale.

\section*{Data availability}

\noindent The \texttt{Chronos} code developed for this work is publicly available at {\small \url{https://github.com/luisgonzalezramirez/Chronos}}. The data underlying this article is available through the CDS (Strasbourg astronomical Data Center), via anonymous ftp to {\small \url{cdsarc.u-strasbg.fr}  (130.79.128.5)} or via {\small \url{http://cdsweb.u-strasbg.fr/cgi-bin/qcat?J/A+A/vol/number}}.

\begin{acknowledgements}
LGR and DB have been funded by grants PID2023-150468NB-I00 by the Spanish Ministry of Science and Innovation/State Agency of Research MCIN/AEI/10.13039/501100011033.  AB acknowledges financial support from project FEDER-UCA-2024-A2-35 funded by Programa Operativo FEDER Andalucía 2021-2027 and by Consejería de Universidad, Investigación e Innovación de la Junta de Andalucía. JO and LS acknowledge financial support from project PID2022-142707NA-I00. “La publicación es parte del proyecto PID2022-142707NA-I00, financiado por MCIN/AEI/10.13039/501100011033/FEDER, UE”.

This work was performed using the \texttt{Python} $\geq 3.8$ scientific ecosystem, including the packages \texttt{NumPy} \citep{harris2020numpy}, \texttt{SciPy} \citep{virtanen2020scipy}, \texttt{pandas} \citep{mckinney2010pandas}, and \texttt{Matplotlib} \citep{hunter2007matplotlib}. Machine learning components rely on \texttt{scikit-learn} \citep{pedregosa2011sklearn} and \texttt{Keras} \citep{chollet2015keras}, while visualization utilities include \texttt{seaborn} \citep{waskom2021seaborn}.

Catalog manipulation and visual inspection were performed using \texttt{TOPCAT} \citep{taylor2005topcat}.

\end{acknowledgements}

\bibliographystyle{aa}
\bibliography{bibliography}

\begin{appendix}

\nolinenumbers

\section{Probability distributions}\label{ap:distributions}

\noindent In this appendix we summarise the probability distributions employed in the \textsc{Chronos} Bayesian hierarchical model, both for priors and latent-variable modelling. Throughout this work, a random variable $X$ with probability density function (PDF) $p(x)$ is denoted as $X \sim \mathcal{D}(\cdot)$, where $\mathcal{D}$ represents the corresponding distribution family. The following families of distributions are used:
\begin{enumerate}[label=(\roman*), topsep=0.2em, partopsep=0pt, itemsep=0.4em, parsep=0pt]
    \item Uniform distribution. The uniform distribution $\mathcal{U}(a,b)$ is defined by the PDF via
    \begin{equation}
    p(x) =
    \begin{cases}
    \dfrac{1}{b-a}, & a \leq x \leq b, \\
    0, & \text{otherwise},
    \end{cases}
    \end{equation}
    with mean $(a+b)/2$ and variance $(b-a)^2/12$. Uniform priors are used for parameters for which only broad physical bounds are known, such as stellar age or effective temperatures.
    \item Normal distribution. The normal distribution $\mathcal{N}(\mu,\sigma^2)$ has a PDF that is expressed via
    \begin{equation}
    p(x) = \dfrac{1}{\sqrt{2\pi\sigma^2}}
    \exp\!\left[-\dfrac{(x-\mu)^2}{2\sigma^2}\right],
    \end{equation}
    with mean $\mu$ and variance $\sigma^2$.

    \item Skew-Normal distribution. $X \sim \mathcal{SN}(\mu,\sigma^2,\alpha)$ extends the normal distribution by including a skewness parameter $\alpha$. Its PDF is
    \begin{equation}
    p(x)=
    \dfrac{2}{\sigma}
    \phi\!\left(\dfrac{x-\mu}{\sigma}\right)
    \Phi\!\left(\alpha\,\dfrac{x-\mu}{\sigma}\right),
    \end{equation}
    where $\phi(\cdot)$ and $\Phi(\cdot)$ denote the standard normal PDF and cumulative distribution function, respectively. For $\alpha=0$, the distribution reduces to a normal distribution.

    \item Gamma distribution. Positive-definite parameters are modelled using the Gamma distribution $X \sim \Gamma(\alpha,\beta)$, where $\alpha$ is the shape parameter and $\beta$ is the rate parameter. The PDF is
    \begin{equation}
    p(x) = \dfrac{\beta^{\alpha}}{\Gamma(\alpha)} x^{\alpha-1} e^{-\beta x},
    \qquad x > 0
    ,\end{equation}
    where the mean, mode and variance are given by $\mathbb{E}[X]=\alpha/\beta$, $\text{mode} = (\alpha - 1)/\beta$ and $\mathrm{Var}(X)=\alpha/\beta^2$, and $\Gamma(x)$ is the Gamma function\footnote{The Gamma function $\Gamma (z)$ is a mathematical function defined as $\Gamma(z)=\int_{0}^{\infty }t^{z-1}e^{-t}dt$, while the Gamma distribution is a continuous probability distribution with $\Gamma(\alpha)$ in its normalization constant.}. Gamma priors ensure positivity while allowing for a wide range of physically plausible values.
    
    \item Beta distribution. Probabilities constrained to $[0,1]$ are assigned Beta distributions $X \sim \mathrm{B}(\alpha,\beta)$ with a PDF,
    \begin{equation}
    p(x) = \dfrac{\Gamma(\alpha+\beta)}{\Gamma(\alpha)\Gamma(\beta)}
    x^{\alpha-1}(1-x)^{\beta-1},\qquad 0 < x < 1,
    \end{equation}
    where the mean, mode and variance are given by $\mathbb{E}[X]=\alpha/(\alpha + \beta)$, $\text{mode} = (\alpha - 1)/(\alpha + \beta - 2)$ and $\mathrm{Var}(X)=\alpha\beta /[(\alpha + \beta)^2(\alpha + \beta + 1)]$. A Beta prior naturally enforces bounded and flexible prior shapes, making it suitable for parameters such as mixture fractions.
    
\end{enumerate}

These distributions are selected to ensure boundedness or positivity when required, while remaining weakly informative so that posterior inference is primarily driven by the data. Together, these prior choices provide a consistent and physically interpretable model that regularises the inference without imposing strong constraints on the posterior distributions.

\section{Spectral regime transition priors}\label{ap:w_spec}

\noindent We describe the construction of the informative prior adopted for the spectral-regime transition temperature $T_0$ in \textsc{Chronos} (implemented in the \texttt{LDB\_threshold} routine). The prior is derived from the geometry of the lithium depletion boundary (LDB) predicted by the NN, so that the transition temperature is informed by physical meaningful regions of the model rather than being treated as a free nuisance parameter.

\subsection{Extraction of the LDB valley geometry}

\begin{figure}[ht]
  
  \includegraphics[width=\columnwidth]{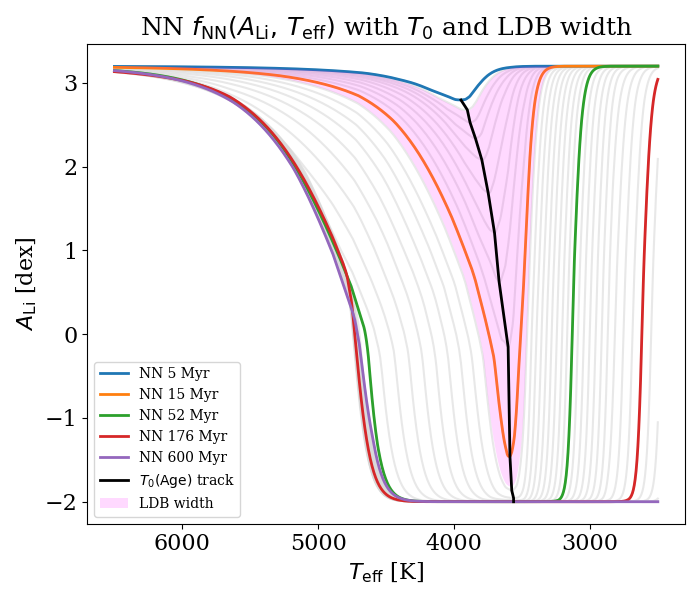}
  \caption{NN predictions of $A_{\rm Li}$ as a function of effective temperature for a grid of stellar ages. The black curve traces the location of the LDB valley minimum per age $T_0({\rm Age})$, while the shaded region illustrates the characteristic width of the valley used to set the kernel scale in the $T_0$ prior. Ages for which the curves approach the depleted plateau contribute progressively less information and are down-weighted.}
  \label{fig:app_LDB_valley}
  \includegraphics[width=\columnwidth]{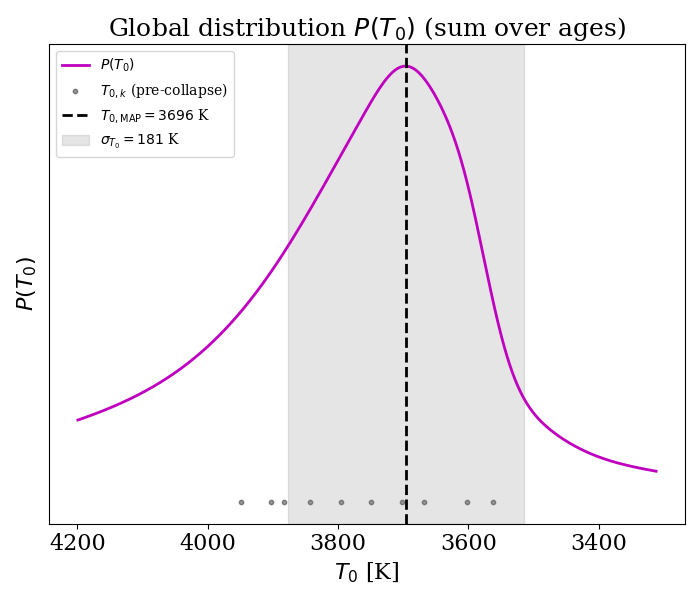}
  \caption{Global distribution $P(T_0)$ obtained from a weighted mixture of age-dependent Gaussian kernels. Grey points indicate the individual $T_{0,k}$ values extracted from pre-collapse ages. The dashed vertical line marks the mode of the distribution, $T_{0,\rm mode}=3696$ K, adopted as the central value of the prior, while the shaded region denotes $2\sigma_{T_0}$.}
  \label{fig:app_PT0}
\end{figure}

\noindent For a grid of stellar ages, $\mathrm{Age}_k$, we evaluate the NN-predicted lithium curves $A_{\rm Li}(T_{\rm eff};\mathrm{Age}_k)$ over a dense effective temperature grid. For each $\mathrm{Age}_k$, we characterise the LDB valley through two geometric quantities:
\begin{enumerate}[label=(\roman*), topsep=0.2em, partopsep=0pt, itemsep=0.4em, parsep=0pt]
    \item The effective temperature of the minimum lithium abundance $T_{0,k} \equiv \arg\min_{T_{\rm eff}}\, A_{\rm Li}(T_{\rm eff};\mathrm{Age}_k)$.
    \item The characteristic valley width ${\rm width}_k \equiv T_{\rm right}-T_{\rm left}$, defined by the temperature interval over which $A_{\rm Li}(T_{\rm left/right};\mathrm{Age}_k)$ equals $A_{\rm Li, min}(\mathrm{Age}_k) + \Delta_{\rm w}$, with an offset of $\Delta_{\rm w}=0.5$ dex above the minimum. This value is chosen to trace the steep transition region of the LDB (Fig. \ref{fig:app_LDB_valley}), capturing the width of the depleted minimum while avoiding both the Li-rich plateau and the fully saturated regime. It also ensures that young ages, where the valley is intrinsically narrow, are properly represented. Here $T_{\rm right}$ and $T_{\rm left}$ are the two temperatures bracketing the minimum that satisfy the above condition, obtained by linear interpolation on the dense $T_{\rm eff}$ grid.
\end{enumerate}

At sufficiently old ages the NN predicted curves approach the minimum-lithium plateau, and the LDB valley becomes weakly defined or fully collapsed for temperatures higher than 4000 K. In this regime, the location of the minimum no longer provides meaningful information about the transition temperature. Such ages are therefore strongly down-weighted in the construction of the global $T_0$ prior, because the location of the minimum becomes increasingly degenerate once the curve approaches the depleted plateau. Figure \ref{fig:app_LDB_valley} illustrates the extraction of the LDB valley geometry from the NN $A_{\rm Li}$ curves, including the age-dependent $T_0({\rm Age})$ track and the associated valley width.

\subsection{\texorpdfstring{Construction of the global $P(T_0)$}{Construction of the global P(T0)}}

\noindent We construct a one-dimensional distribution $P(T_0)$ by combining the age-dependent transition temperatures $\{T_{0,k}\}$ into a weighted Gaussian mixture:
\begin{equation}
P(T_0)=\frac{1}{\mathcal Z}
\sum_{k=1}^{K}
w_k\,
\mathcal{N}\!\left(T_0 \mid T_{0,k}, \sigma_k^2\right)
\end{equation}
where $K$ is the number of sampled ages and $\mathcal Z$ ensures normalisation. The kernel width is tied to the LDB morphology,
\begin{equation}
\sigma_k=\max\!\left(\mathrm{width}_k/f_w,\,\sigma_{\rm floor}\right),
\end{equation}
so that broader valleys produce wider kernels while a minimum scale $\sigma_{\rm floor}$ prevents numerical collapse.

Because the NN predictions are evaluated on a non-uniform age grid, the mixture is constructed as a Riemann-sum approximation to an integral over age. We therefore weight each term by the local linear-age spacing $\Delta \mathrm{Age}_k \equiv (\mathrm{Age}_{k+1} - \mathrm{Age}_{k-1})/2$ with forward-backward differences at the boundaries, preventing densely sampled regions from dominating the distribution.

\subsection{Information-based weighting and suppression of weakly informative ages}

\noindent The contribution of each age-dependent kernel is modulated by a set of information-based weights designed to ensure that $P(T_0)$ is driven by genuine regions of the LDB morphology.

The exponential down-weighting is given by
\begin{equation}\label{eq:weight_age}
    w_{k,\rm age} = \exp{\left[-\dfrac{\log_{10}{(\mathrm{Age}_k)} - \log_{10}{(\mathrm{Age}_{\rm min})}}{\tau_{\rm log}}\right]}
.\end{equation}

\noindent These information-based weights for each kernel reflect the diagnostic power of the LDB morphology. The total weight is written schematically as
\begin{equation}\label{eq:w_k_age}
w_k \propto
\Delta \mathrm{Age}_k
\times
w_{k,\rm geom}
\times
w_{k,\rm age}
\times
w_{k,\rm track}
\end{equation}
with the following components:
\begin{enumerate}[label=(\roman*), topsep=0.2em, partopsep=0pt, itemsep=0.4em, parsep=0pt]
    \item Geometric weight:
    \[
    w_{k,\rm geom}=
    \left(\frac{\mathrm{width}_k}{\mathrm{med}(\mathrm{width})}\right)\times
    \left(\frac{\mathrm{depth}_k}{\mathrm{med}(\mathrm{depth})}\right)
    \]
    so that ages with wide and deep LDB valleys contribute more strongly than those with shallow or poorly defined transitions.
    \item Age preference: given by Eq. (\ref{eq:weight_age}), where $\mathrm{Age}_{\min}$ is the youngest valid pre-collapse age. We adopt $\tau_{\rm log}=0.2$, corresponding to an e-folding scale of 0.2 dex (a factor of $\sim$1.6 in age), providing a moderate preference for young ages.
    \item Track-slope weight:
    \[w_{k,\rm track} \propto
    \left|\frac{dT_0}{d\log_{10}(\mathrm{Age})}\right|\]
    which suppresses locally flat segments of the $T_0(\mathrm{Age})$ relation and prevents artificial stacking when neighbouring ages yield nearly identical transition temperatures.
\end{enumerate}

Ages for which the LDB valley has collapsed onto the depleted plateau are automatically excluded, as their effective width becomes zero and they carry negligible information.

\subsection{Adopted prior and choice of central value}

\noindent The resulting distribution $P(T_0)$ is generally non-Gaussian and exhibits an asymmetry, reflecting both the intrinsic evolution of the LDB morphology with age and the adopted information-based weighting scheme. Any residual non-Gaussian features arise from the superposition of kernels associated with neighbouring pre-collapse ages and do not indicate the presence of multiple physically distinct transition temperatures. For this reason, we then characterise the central value of the prior by the mode $T_{0,\rm mode}$ of the Gaussian mixture. We obtain $T_{0,\rm mode} \simeq 3696 {\,\rm K}$ with an effective width of $2\sigma_{T_0} \simeq 181 {\,\rm K}$, as half the central 68\% credible interval of $P(T_0)$, mapped onto an equivalent Gaussian $\sigma$. This motivates the informative priors adopted for $T_0$ and $\sigma_{T_0}$, reported in Table \ref{tab:chronos_priors}.

We show in Fig. \ref{fig:app_PT0} the resulting global distribution $P(T_0)$ obtained from the weighted mixture of age-dependent kernels, highlighting the adopted $T_{0,\rm mode}$ and the effective prior width. After correcting for the non-uniform age grid, the inferred $P(T_0)$ is stable against changes in sampling across different ages, indicating that its shape is primarily determined by the NN-predicted $T_0(\mathrm{Age})$ track and by the intrinsic morphology of the LDB valley. In the present case, the dominant contribution arises from ages $\sim$5--20 Myr, where the LDB is both well defined and rapidly evolving in effective temperature.

\section{MLP training and validation}\label{ap:mlp}

\noindent We describe the input-output scaling process and the training procedure of the multilayer perceptron (MLP), a type of neural network (NN), used in \textsc{Chronos} to emulate the BT-Settl lithium depletion predictions. The goal of the network is to provide a fast, smooth approximation to the model mean lithium abundance given by $\mu_{A_{\rm Li}} = f_{\rm NN}(\mathrm{Age}, T_{\rm eff})$, over the age and effective temperature domain relevant to the LDB analysis.

\subsection{Input representation and normalisation}\label{ap:mlp:input}

\noindent For a sample of $N$ sources, the network input is defined as
\begin{equation}
\mathbf{X}_{\rm input} =
\begin{pmatrix}
\mathrm{Age} & T_{{\rm eff,true},1} \\
\vdots & \vdots \\
\mathrm{Age} & T_{{\rm eff,true},N}
\end{pmatrix}
\in \mathbb{R}^{N \times 2}
,\end{equation}
where the cluster age is repeated for all sources.

To improve numerical conditioning and reduce feature skewness, both input variables are transformed using a Box-Cox power transform:
\begin{equation}
x_{ij} =
\begin{cases}
\dfrac{X_{{\rm input},ij}^{\lambda_j}-1}{\lambda_j}, & \lambda_j \neq 0, \\[6pt]
\log(X_{{\rm input},ij}), & \lambda_j = 0,
\end{cases}
\end{equation}
where $\lambda_j$ is the Box-Cox parameter associated with feature $j$ (age or effective temperature). The transformed inputs are then linearly rescaled to the interval $[0,1]$ using a min–max normalisation.

\subsection{Training dataset and target conditioning}\label{ap:mlp:training}

\noindent The NN is trained on a precomputed grid of BT-Settl evolutionary models \citep{Allard2012} with solar metallicity, tabulated as triples $\{\mathrm{Age}\,[{\rm Myr}],\; T_{\rm eff}\,[{\rm K}],\; A_{\rm Li}\,[{\rm dex}]\}$. To match the domain explored during inference, the training set is restricted to ages in the range $5 \le \mathrm{Age} \le 600$ Myr.

The raw BT-Settl tables may contain fully depleted lithium fraction values Li/Li$_0 = 0$ translated into $A_{\rm Li} = -\infty$. For numerical stability, these values are replaced by a finite lower bound and the target is subsequently clipped to the physical interval $A_{\rm Li}\in[A_{\rm Li,min},A_{\rm Li,max}]$, with $A_{\rm Li,min}=-2.0$ dex and $A_{\rm Li,max}=3.2$ dex. The clipped target is mapped to the $[0,\,1]$ interval as
\begin{equation}
A_{\rm Li}^{\rm scaled}=
\frac{\mathrm{clip}(A_{\rm Li},A_{\rm Li,min},A_{\rm Li,max})-A_{\rm Li,min}}
{A_{\rm Li,max}-A_{\rm Li,min}}
.\end{equation}

The NN output layer uses a sigmoid activation to ensure that predictions remain within $[0,1]$ by construction. Physical $A_{\rm Li}$ values are recovered through the inverse affine transformation.

\subsection{Network architecture and optimisation}\label{ap:mlp:arch}

\noindent The NN is a fully connected multilayer perceptron with four hidden layers of 64 neurons each and rectified linear unit (ReLU) activations, followed by a single sigmoid output neuron. 

Let $\mathbf{X}^{\rm scaled}\in\mathbb{R}^{N\times 2}$. The forward propagation is
\setlength{\jot}{2pt}
\begin{align}
\mathbf{Z}^{(1)} &= \mathbf{X}^{\rm scaled}\mathbf{W}_1 + \mathbf{b}_1,
& \mathbf{A}^{(1)} &= \mathrm{ReLU}\!\left(\mathbf{Z}^{(1)}\right), \\
\mathbf{Z}^{(k)} &= \mathbf{A}^{(k-1)}\mathbf{W}_k + \mathbf{b}_k,
& \mathbf{A}^{(k)} &= \mathrm{ReLU}\!\left(\mathbf{Z}^{(k)}\right), \\
\mathbf{Z}^{(5)} &= \mathbf{A}^{(4)}\mathbf{W}_5 + \mathbf{b}_5,
& \hat{\mathbf{y}} &= \mathrm{sigmoid}(\mathbf{Z}^{(5)}),
\end{align}
where $\mathbf{W}_k \in \mathbb{R}^{n_{k-1}\times n_k}$ and $\mathbf{b}_k\in\mathbb{R}^{n_k}$, $n_0=2$, $n_1,n_2,n_3,n_4=64$ and $n_5=1$.

In the implementation used in this work, training was performed using the Adam optimiser with a learning rate of $10^{-3}$, a batch size of 32, and 400 epochs. An explicit validation split of 20\% of the training grid was adopted to monitor out-of-sample performance during optimisation.

We also minimised binary cross-entropy as a convenient loss for a target constrained to $[0,1]$ via a sigmoid output; the physical variable is recovered by an affine inverse mapping. The root-mean-square error (RMSE) in the scaled target space is monitored for both training and validation subsets as an auxiliary diagnostic.

\begin{figure}[!ht]

    \includegraphics[width=0.475\textwidth]{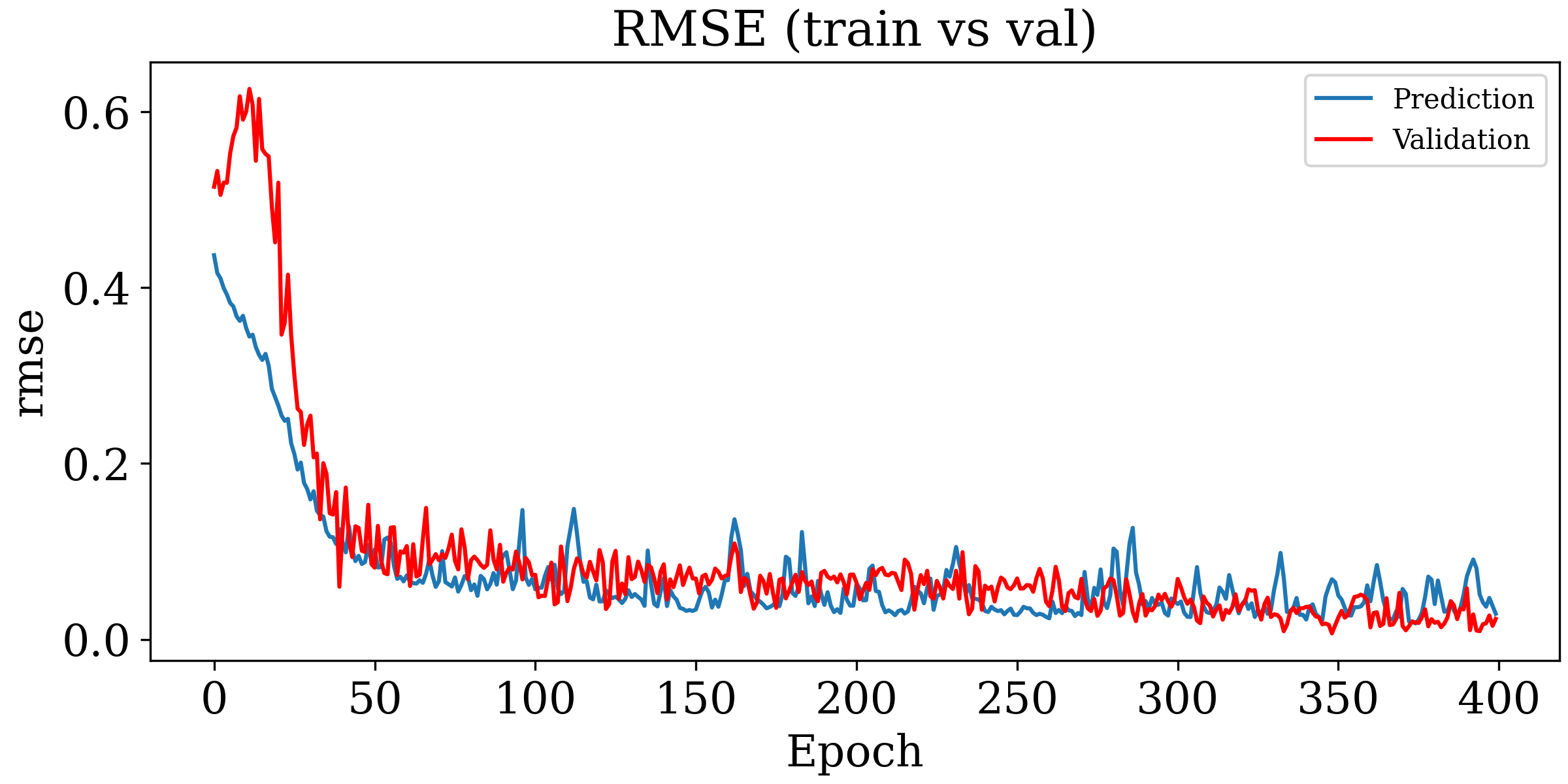}
    \caption{Training diagnostics of the MLP: root-mean-square error (RMSE) in the scaled target space for training (blue) and validation (red) subsets. Both curves stabilise after an initial rapid decrease.}
    \label{fig:mlp_loss_rmse}
\end{figure}

\begin{figure}[!ht]
    \centering
    \includegraphics[width=0.395\textwidth]{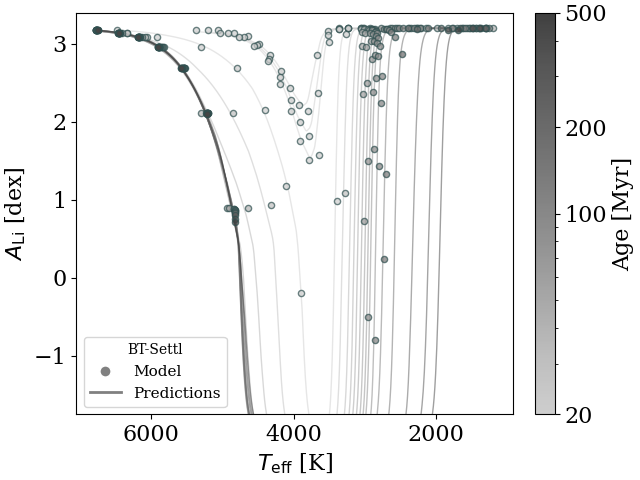}
    \caption{MLP predictions (lines) compared to BT-Settl tabulated values (points) across the set of training ages. Curves are colour-coded by age. The NN reproduces the overall morphology of lithium depletion and the sharp LDB transition.}
    \label{fig:mlp_predictions_all}
    \includegraphics[width=0.395\textwidth]{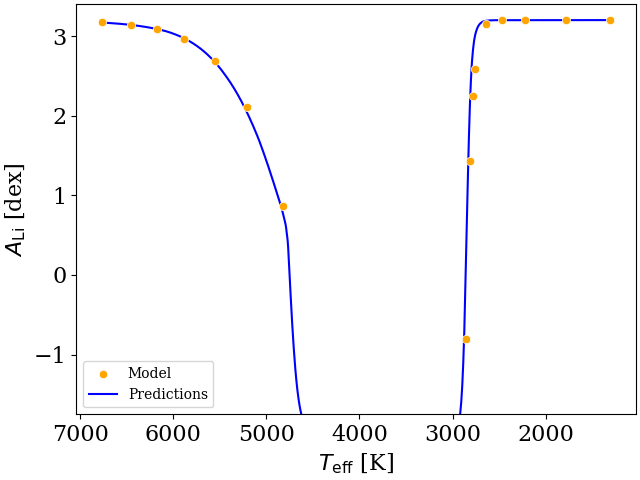}
    \caption{Example of NN predictions fidelity at $\mathrm{Age}=120$ Myr. Points show BT-Settl tabulated $A_{\rm Li}$ and the solid line shows the MLP prediction evaluated on a dense $T_{\rm eff}$ grid.}
    \label{fig:mlp_predictions_120}
\end{figure}

\subsection{Convergence diagnostics and fidelity to BT-Settl models}\label{ap:mlp:pred_loss}

\noindent To quantify the fidelity of the neural network, we compare its predictions directly with the original BT-Settl grid across the full training domain. The root-mean-square error (RMSE) in the scaled target space is found to be $\sim$0.03--0.05 for both training and validation subsets (Fig. \ref{fig:mlp_loss_rmse}). Converting to physical units, this corresponds to an error of $\sim$0.15--0.25 dex in $A_{\rm Li}$, given the adopted scaling range.

The neural network approximation reproduces the overall morphology of the lithium depletion curves, including the plateau, depletion onset, and sharp transition at the LDB (Fig. \ref{fig:mlp_predictions_all}). We note that this level of approximation error is comparable to or smaller than the observational uncertainties and intrinsic dispersion terms included in the hierarchical model, and therefore does not dominate the inference. This behaviour indicates stable optimisation without significant overfitting. Figure \ref{fig:mlp_predictions_all} illustrates the agreement across ages: the MLP reproduces the lithium plateau, the onset of depletion, and the sharp transition at the lithium depletion boundary.

As an illustrative example, Fig. \ref{fig:mlp_predictions_120} shows the predicted curve at $\mathrm{Age}=120$ Myr together with the BT-Settl tabulated values at that age, demonstrating that the NN captures both the smooth FGK depletion trend and the steep transition towards fully depleted ultracool dwarfs.

Finally, the trained neural network is exported as a lightweight object containing the NN and the fitted input scalers. This ensures that the mapping $(\mathrm{Age},T_{\rm eff}) \mapsto \mu_{A_{\rm Li}}$ is deterministic and consistent across training, synthetic-data generation, and Bayesian inference within \textsc{Chronos}.

\section{Synthetic data generation}\label{ap:fake_data}

\noindent To validate \textsc{Chronos}, we perform posterior predictive checks (PPCs) and we quantify the sensitivity of LDB age inference to sample composition. To this end, we implemented a synthetic (i.e. fake) data generator. The generator follows the same forward-modelling assumptions adopted for inference (Sect. \ref{sc:bhm}), while enabling controlled experiments in which the number of cool objects ($T_{\rm eff}<3700\,{\rm K}$) is fixed.

\subsection{Global and local draws}

\noindent A synthetic dataset contains $N$ objects, indexed by $i=1,\dots,N$. In the configuration adopted in this work, the central cluster age is fixed to the input value $\mathrm{Age}_{\rm input}$, corresponding to the mean of the age prior used for the experiment $\mathrm{Age}_{\rm LDB} \equiv \mathrm{Age}_{\rm input}$. An intrinsic age dispersion $\sigma_{\rm Age}$ is introduced to mimic a finite spread around the central age:
\begin{equation}\label{eq:fake_age}
\mathrm{Age}_i \sim \mathcal{N}\left(\mathrm{Age}_{\rm LDB},\sigma_{\rm Age}^{2}\right)
.\end{equation}

Rather than explicitly drawing $A_{\rm Li}$ from individual stellar ages, we generate a coeval mean lithium pattern at $\mathrm{Age}_{\rm LDB}$ and propagate the age dispersion into lithium space through the local derivative of the neural network model (see Sect. \ref{ap:fake_data:ali_true}).

For the FGK–UCD/M transition, the regime parameters are drawn once per dataset from their priors $T_0 \sim p(T_0)$, $\sigma_{T_0} \sim p(\sigma_{T_0})$ and we define the smooth FGK membership weight $\omega_{\mathrm{FGK},i}$ via the logistic transition described in Sect. \ref{sc:bhm} (Eq. \ref{eq:sigmoid}).

\subsection{Enforcing the UCD/M-dwarf content}

\noindent To enforce exactly $n_{\rm below}$ objects with $T_{\rm eff}<3700\,{\rm K}$, we draw $N-n_{\rm below}$ effective temperatures from the adopted $T_{\rm eff}$ prior truncated to $[3700,T_{\rm eff,max}]$. The remaining $n_{\rm below}$ temperatures are generated in the UCD/M regime using conditionally on the sampled per-source ages $\mathrm{Age}_i$. For UCD/M objects, we first estimate the effective temperature, $T_{\rm sat}(\mathrm{Age}_i)$, at which the NN lithium prediction reaches the saturation level. We then draw either (i) a saturated object with $T_{\rm eff}<T_{\rm sat}$ with a small probability, or (ii) a depleted object with $T_{\rm eff}$ concentrated near the vertical LDB wall defined by $\mu_{A_{\rm Li}}(\mathrm{Age}_i,T_{\rm mid} \pm \Delta T_{\rm eff})$, where $T_{\rm mid}$ is the midpoint of the LDB wall and $\Delta T_{\rm eff}$ is the width between $T_{\rm eff}$ on the bottom of the vertical wall and $T_{\rm sat}$.

This procedure produces a realistic distribution of cool objects around the LDB while keeping the number of objects below 3700 K fixed (see Appendix \ref{ap:w_spec}).

\subsection{Lithium abundance generation}\label{ap:fake_data:ali_true}

\noindent For each source, the reference lithium abundance is obtained by evaluating the trained neural network at the central cluster age given by $\mu_{A_{\mathrm{Li}},i} \equiv f_{\mathrm{NN}}(\mathrm{Age}_{\rm LDB},T_{\mathrm{eff,true},i})$. Baseline (i.e. quiet) lithium abundances are generated by adding intrinsic dispersion with a regime-dependent scale $\sigma_{A_{\rm Li},\mathrm{intr}}(T_{\rm eff})$ defined by the model specification.

To encode an intrinsic age spread without introducing explicit age-dependent shifts in the mean lithium pattern, the age dispersion is propagated into lithium space through the neural network using a local finite-difference approximation:
\begin{equation}
\sigma_{A_{\rm Li},\mathrm{age},i} =
\left|
\frac{\partial \mu_{A_{\rm Li}}}{\partial \mathrm{Age}}
\right|_{T_{\mathrm{eff,true},i}}\times\,
\sigma_{\rm Age}
\end{equation}

The effective lithium dispersion used in the synthetic generator is then given by the quadratic sum of intrinsic $A_{\rm Li}$ dispersion and the broadening induced by an age spread described above, while preserving a strictly coeval mean trend at $\mathrm{Age}_{\rm LDB}$.

\subsubsection{Rotators in synthetic data}\label{ap:fake_data:rotators}

\noindent To test the FGK rotating-mixture component, the generator can inject rotators following the same hierarchical mixture family adopted in inference. For each synthetic dataset, the global rotation hyperparameters are drawn once from their priors, including the rotator fraction $p_{\rm rot}$, the enhancement scale $\Delta A_{\rm Li,rot}$, the additional dispersion term $\sigma_{A_{\rm Li},\rm rot}$, and the skewness parameter $\alpha_{\rm FGK}$. These hyperparameters define a coherent rotating subpopulation shared across all FGK stars in a given synthetic realisation.

Consistent with observations, fast rotators are generated only in the FGK regime. The realised number of fast rotating stars is constrained to match the sampled global fraction $p_{\rm rot}$ up to integer rounding, ensuring that the synthetic catalogue reflects the intended mixture proportion while remaining statistically consistent with the hierarchical model. Conditional on ${\rm rot}_i=1$, true lithium abundances are drawn from the rotating component of Eq. (\ref{eq:latent_two_component_mixture}).

To prevent unphysical realisations (e.g. fast rotating stars with lithium below the NN mean prediction or exceeding the saturation plateau), fast rotator draws are generated by rejection sampling from the corresponding Skew-Normal distribution truncated to the physically allowed interval: $\mu_{A_{\rm Li}}(T_{\rm eff}) \le A_{{\rm Li,true},i} \le A_{\rm Li,sat}-\epsilon$ where $\epsilon$ is a small safety margin. This truncation is applied only in the synthetic generator and is not part of the likelihood model used in inference. Optionally, to avoid artificial accumulation of slow rotating FGK stars at the depletion floor in particular realisations, the generator may reassign a small number of sources between components while preserving the total number of rotators. This adjustment does not alter the global mixture fraction.

\subsection{Observational layer and exported catalogue}\label{ap:fake_data:obs}

\noindent Finally, the observed quantities are produced by adding Gaussian errors to the latent values. Observational uncertainties for $A_{\rm Li}$ are taken source-by-source from the reference dataset and applied following Eq. (\ref{eq:likelihood}) for $A_{\rm Li}$ with an analogous prescription for $T_{\rm eff}$ according to the adopted likelihood model. The exported synthetic data stores both observables and auxiliary quantities (e.g. $\mathrm{Age}_i$, $\mathrm{Age}_{\rm LDB}$, ${\rm rot}_i$) to facilitate controlled validation experiments. The synthetic generator is independent of any specific reference dataset and can be applied to arbitrary stellar populations by modifying the adopted priors and neural network model, without altering the underlying model assumptions.

\section{Convergence diagnostics summary}\label{ap:convergence}

\begin{table}[!ht]
\vspace{-10pt}
\centering
\caption{Summary of chains convergence diagnostics.}
\vspace{-7.5pt}
\renewcommand{\arraystretch}{1.5}
\begin{tabular}{l@{\hspace{1.75pt}}c@{\hspace{4.5pt}}l@{\hspace{3.75pt}}r@{\hspace{1.25pt}}l@{\hspace{3.75pt}}l@{\hspace{2.25pt}}l}
\toprule
$\boldsymbol{\Theta}$ & $\hat{R}_{\rm mean}$ & \textbf{ESS} & Mode$_{-\Delta_-}^{+\Delta_+}$ && MCSE& \\
\hline
$\sigma_{T_0}$                & 1.0005 & 42122 & $98.72_{-9.92}^{+10.45}$ &K   & 0.0492 &K \\
$T_{0}$                       & 1.0001 & 42530 & $3697_{-25.04}^{+24.62}$ &K   & 0.1197 &K \\
$\mathrm{Age}_{\rm LDB}$      & 1.0002 & 4697  & $124.53_{-2.70}^{+3.34}$ &Myr & 0.0450 &Myr \\
$\sigma_{A_{\rm Li}}$         & 1.0003 & 8794  & $0.243_{-0.025}^{+0.035}$ &dex & 0.0003 &dex \\
$p_{\rm rot}$                 & 1.0002 & 22798 & $0.322_{-0.043}^{+0.061}$ && 0.0004 & \\
$\Delta A_{\rm Li,rot}$       & 1.0004 & 13542 & $1.78_{-0.438}^{+0.335}$  &dex & 0.0036 &dex \\
$\sigma_{A_{\rm Li},\rm rot}$ & 1.0006 & 15465 & $1.08_{-0.177}^{+0.384}$  &dex & 0.0024 &dex \\
$\alpha_{\rm FGK}$            & 0.9999 & 22165 & $1.06_{-0.392}^{+1.69}$   && 0.0103 & \\
\hline\hline
\end{tabular}
\renewcommand{\arraystretch}{1.0}
\tablefoot{
  Gelman--Rubin $\hat{R}$, effective sample size
  (ESS$_\mathrm{bulk}$) over 4 chains of 15{,}000 steps each, posterior mode with the 95\% highest-density interval (HDI) uncertainty, and MCSE of the mean for each parameter.
}
\label{tab:convergence_global}
\vspace{-7.5pt}
\end{table}

\section{Synthetic data results}\label{ap:fake_data_results}

\begin{table}[!ht]

\caption{Prior distributions used to generate the 20 Myr synthetic dataset and weighted mean of all $n_{T_{\rm eff}<3700\,\mathrm{K}}$ posterior modes ($\mu_{\rm modes}$) with bootstrap error.}
\vspace{-7.5pt}
\begin{tabular}{l@{\hspace{3.75pt}}ll}
\toprule
$\boldsymbol{\Theta}$  & \textbf{Prior distribution} & $\mu_{\rm modes}$ \\
\midrule

\multicolumn{3}{l}{\textbf{Main global parameters}}\\
\midrule

$\mathrm{Age}$ [Myr]
& $\mathcal{N}(\mu=20,\;\sigma=1)$
& $19.59 \pm 0.08$ \\[2pt]

$\sigma_{A_{\rm Li}}$ [dex]
& $\Gamma(\alpha=4.5,\;\beta=30)$
& $0.05328 \pm 0.01199$ \\

\midrule

\multicolumn{3}{l}{\textbf{Global parameters for regime transition} $\theta_r$} \\
\midrule

$T_0$ [K]
& $\mathcal{N}(\mu=3700,\;\sigma=25)$
& $3706.73 \pm 9.80$ \\[2pt]

$\sigma_{T_0}$ [K]
& $\Gamma(\alpha=100,\;\beta=1)$
& $49.40 \pm 0.98$ \\

\midrule

\multicolumn{3}{l}{\textbf{Global parameters for FGK rotation} $\theta_{\rm rot}$} \\
\midrule

$p_{\rm rot}$ 
& $\mathrm{B}(2,25)$
& $0.153 \pm 0.012$ \\[2pt]

$\Delta A_{\rm Li, rot}$ [dex]
& $\Gamma(\alpha=3,\;\beta=3)$
& $0.919 \pm 0.050$ \\[2pt]

$\sigma_{A_{\rm Li},{\rm rot}}$ [dex]
& $\Gamma(\alpha=1,\;\beta=50)$
& $0.003 \pm 0.001$ \\[2pt]

$\alpha_{\rm FGK}$ 
& $\Gamma(\alpha=12,\;\beta=50)$
& $0.256 \pm 0.015$ \\

\midrule
\bottomrule
\end{tabular}

\end{table}

\begin{table}[t!]
\vspace{-3.5pt}
\caption{Summary of 120 Myr synthetic dataset.}
\vspace{-7.5pt}
\begin{tabular}{l@{\hspace{3.75pt}}ll}
\toprule
$\boldsymbol{\Theta}$  & \textbf{Prior distribution} & $\mu_{\rm modes}$ \\
\midrule

\multicolumn{3}{l}{\textbf{Main global parameters}} \\
\midrule

$\mathrm{Age}$ [Myr]
& $\mathcal{N}(\mu=120,\;\sigma=6)$
& $120.78 \pm 0.57$ \\[2pt]

$\sigma_{A_{\rm Li}}$ [dex]
& $\Gamma(\alpha=4.5,\;\beta=30)$
& $0.01900 \pm 0.00221$ \\

\midrule

\multicolumn{3}{l}{\textbf{Global parameters for regime transition} $\theta_r$}  \\
\midrule

$T_0$ [K]
& $\mathcal{N}(\mu=3700,\;\sigma=25)$
& $3698.50 \pm 6.05$ \\[2pt]

$\sigma_{T_0}$ [K]
& $\Gamma(\alpha=100,\;\beta=1)$
& $46.93 \pm 3.20$ \\

\midrule

\multicolumn{3}{l}{\textbf{Global parameters for FGK rotation} $\theta_{\rm rot}$}  \\
\midrule

$p_{\rm rot}$ 
& $\mathrm{B}(4,10)$
& $0.373 \pm 0.017$ \\[2pt]

$\Delta A_{\rm Li, rot}$ [dex]
& $\Gamma(\alpha=3,\;\beta=3)$
& $1.520 \pm 0.115$ \\[2pt]

$\sigma_{A_{\rm Li},{\rm rot}}$ [dex]
& $\Gamma(\alpha=3,\;\beta=3)$
& $1.306 \pm 0.099$ \\[2pt]

$\alpha_{\rm FGK}$ 
& $\Gamma(\alpha=12,\;\beta=50)$
& $0.313 \pm 0.019$ \\

\midrule
\bottomrule
\end{tabular}
\end{table}

\begin{table}[t!]
\vspace{-3.5pt}
\caption{Summary of 400 Myr synthetic dataset.}
\vspace{-7.5pt}

\begin{tabular}{l@{\hspace{3.75pt}}ll}
\toprule
$\boldsymbol{\Theta}$  & \textbf{Prior distribution} & $\mu_{\rm modes}$ \\
\midrule

\multicolumn{3}{l}{\textbf{Main global parameters}} \\
\midrule

$\mathrm{Age}$ [Myr]
& $\mathcal{N}(\mu=400,\;\sigma=20)$
& $399.35 \pm 1.40$ \\[2pt]

$\sigma_{A_{\rm Li}}$ [dex]
& $\Gamma(\alpha=4.5,\;\beta=30)$
& $0.01687 \pm 0.00210$ \\

\midrule

\multicolumn{3}{l}{\textbf{Global parameters for regime transition} $\theta_r$}  \\
\midrule

$T_0$ [K]
& $\mathcal{N}(\mu=3700,\;\sigma=25)$
& $3694.69 \pm 6.90$ \\[2pt]

$\sigma_{T_0}$ [K]
& $\Gamma(\alpha=100,\;\beta=1)$
& $48.78 \pm 1.56$ \\

\midrule

\multicolumn{3}{l}{\textbf{Global parameters for FGK rotation} $\theta_{\rm rot}$}  \\
\midrule

$p_{\rm rot}$ 
& $\mathrm{B}(4,10)$
& $0.320 \pm 0.009$ \\[2pt]

$\Delta A_{\rm Li, rot}$ [dex]
& $\Gamma(\alpha=3,\;\beta=3)$
& $1.750 \pm 0.149$ \\[2pt]

$\sigma_{A_{\rm Li},{\rm rot}}$ [dex]
& $\Gamma(\alpha=3,\;\beta=3)$
& $1.312 \pm 0.050$ \\[2pt]

$\alpha_{\rm FGK}$ 
& $\Gamma(\alpha=12,\;\beta=50)$
& $0.218 \pm 0.016$ \\

\midrule
\bottomrule
\end{tabular}
\vspace{-7.5pt}
\tablefoot{The prior distributions are displayed in Table \ref{tab:chronos_priors} with only one change on $\sigma_{A_{\rm Li}, \rm rot}$ where we use $\Gamma(\alpha=3,\:\beta=2)$ for sampling efficiency.}
\vspace{-12pt}
\end{table}

\noindent In this appendix we report the posterior results for synthetic datasets generated from known priors. The inferred global parameters and ages are consistent with the injected values, with stable MCMC convergence in all cases. For each MCMC run $R$, we identify the mode as the draw that maximises the sampled log-posterior (lp) over all chains. For each parameter, we then extract its value at this mode draw, yielding one scalar estimate per run. The reported mean of posterior modes ($\mu_{\rm modes}$) is simply the arithmetic average of these mode estimates across runs:
\begin{equation}
    \mu_{\rm modes} = \frac{1}{R}\sum^R_{r=1} \theta^{(r)}_{\rm mode},
\end{equation}
where $\theta^{(r)}_{\rm mode}$ is the mode estimate of parameter $\theta$.

\section{Residual PIT summary on Pleiades}\label{ap:residuals_z}

\noindent In this appendix, we present a complementary calibration diagnostic based on posterior predictive probability integral transform (PIT) values. These residuals differ from the measurement-model residuals discussed in Sect. \ref{sc:residuals}, as they probe the calibration of the full posterior predictive distribution rather than the Gaussian observational likelihood alone. For each source $i$ and observable $j$, we define:
\begin{equation}
{\rm PIT}_{i,j} \equiv 
\Pr\!\left(y_{i,j}^{\rm rep} \le y_{i,j}^{\rm obs} \mid \mathcal{D}\right),
\quad
z_{i,j} \equiv \Phi^{-1}({\rm PIT}_{i,j})
\end{equation}

In this notation, we use ${\rm rep}$ to denote a replicated observation drawn from the posterior predictive distribution and $\Phi^{-1}$ is the inverse of the standard normal CDF. Under correct predictive calibration, the PIT values are uniformly distributed on $[0,1]$, and the transformed residuals $z_{i,j}$ follow approximately a $\mathcal N(0,1)$ distribution.

\begin{table}[!ht]

\caption{PIT predictive residual statistics for the observables.}
\label{tab:residual_stats}
\renewcommand{\arraystretch}{1.3}
\begin{tabular}{lcccccc}
\toprule
\textbf{Obs.} 
& $\langle z\rangle$ 
& $\sigma(z)$ 
& $N$ 
& $|z|\le1$ 
& $|z|\le2$ 
& $|z|>3$ \\
\hline
$A_{\rm Li}$ 
& $0.02$ 
& $0.17$ 
& $122$ 
& $99.2\%$ 
& $100.0\%$ 
& $0$ \\
$T_{\rm eff}$ 
& $0.46$ 
& $2.83$ 
& $122$ 
& $88.5\%$ 
& $93.4\%$ 
& $6$ \\
\hline\hline
\end{tabular}
\tablefoot{The quoted fractions indicate the proportion of sources within the corresponding residual thresholds. For $T_{\rm eff}$, the large standard deviation is driven by some extreme outliers (see Table \ref{tab:teff_outliers}).}
\end{table}

The statistics reported in Table \ref{tab:residual_stats} are computed star-wise from the predictive residuals $z_{i,j}$ and should be interpreted as a global predictive calibration diagnostic. The most extreme outliers correspond to hot stars ($T_{\rm eff}\gtrsim7000\,\mathrm{K}$) that lie outside the calibrated temperature range of the NN surrogate (see Table \ref{tab:teff_outliers}). Their large predictive residuals reflect extrapolation of the physical NN rather than failure of the Gaussian measurement model in the LDB-sensitive regime.

\begin{table}[!ht]

\caption{Stars with the largest residuals in $T_{\rm eff}$.}
\label{tab:teff_outliers}
\renewcommand{\arraystretch}{1.3}
\begin{tabular}{cccccc}
\toprule
$T_{\rm eff,obs}$ [K] 
& $T_{\rm eff,pred}$ [K] 
& $\Delta T_{\rm eff}$ [K] 
& $r_{T_{\rm eff}}$ 
& $|r_{T_{\rm eff}}|$ \\
\hline
8069 & 6998 & $+1071$ & $+21.43$ & 21.43 \\
7669 & 6996 & $+673$  & $+13.45$ & 13.45 \\
7554 & 6996 & $+558$  & $+11.17$ & 11.17 \\
7535 & 6995 & $+540$  & $+10.79$ & 10.79 \\
7427 & 6994 & $+433$  & $+8.65$  & 8.65 \\
2641 & 2810 & $-169$  & $-3.39$  & 3.39 \\
\hline\hline
\end{tabular}
\tablefoot{Only 6 out of 122 sources have $|r_{T_{\rm eff}}|>3$. The most extreme outliers correspond to stars with $T_{\rm eff}\gtrsim7000$ K, rounded to the nearest unit.}
\vspace{-10pt}
\end{table}

\section{Consistency check of PSIS-LOO re-weighting}\label{ap:loo_sanity}

\noindent Because PSIS-LOO in hierarchical mixture models may yield unstable importance weights \citep{Vehtari2016}, we performed an internal consistency check of the reweighting implementation. This diagnostic is constructed on the global parameter $\mathrm{Age}_{\rm LDB}$ and should not be interpreted as a test of predictive calibration, since $\mathrm{Age}_{\rm LDB}$ is not an observed quantity. 

\begin{figure}[!ht]
    \includegraphics[page=1,width=\columnwidth, trim=0cm 0cm 0cm 0cm, clip]{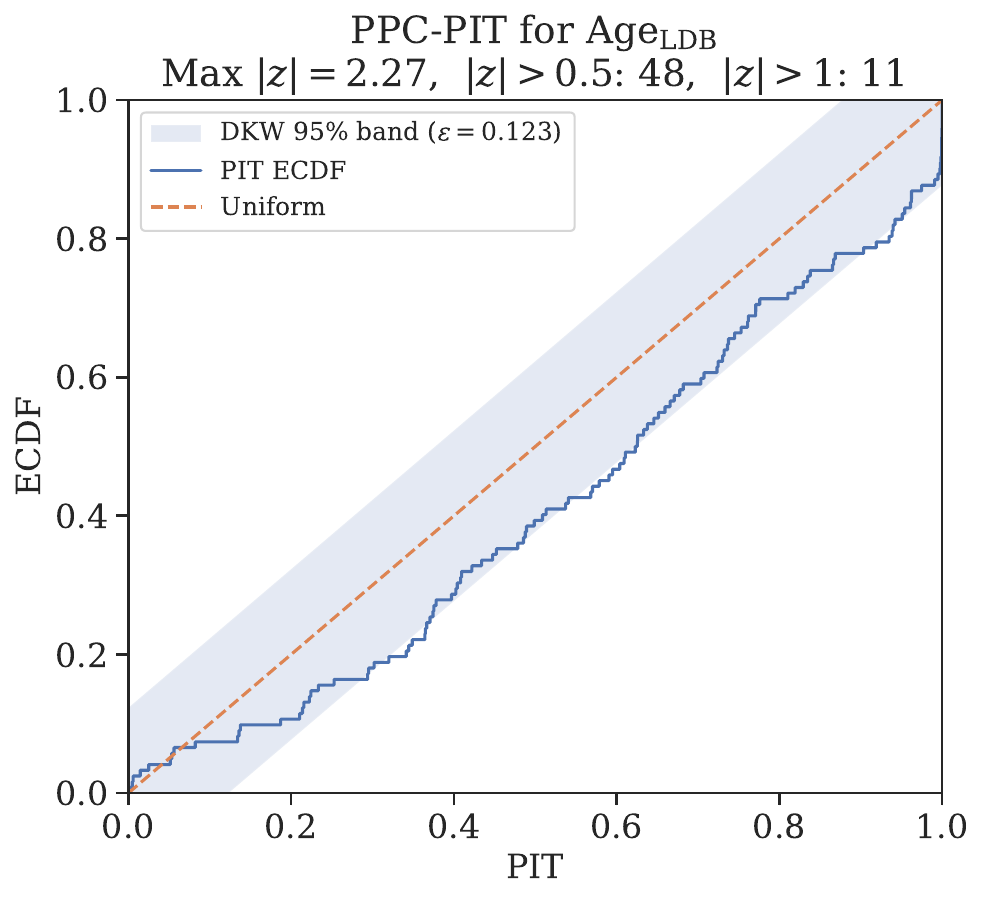}
    \caption{ECDF of PPC–PIT values for $\mathrm{Age}_{\rm LDB}$ compared to the uniform reference (dashed line). The shaded region shows the 95\% DKW confidence band.}
    \label{fig:ECDF_DKW}
\end{figure}

For each source $i$, we draw a replicated age ${\rm Age}^{\rm rep}$ from the LOO-reweighted posterior $p({\rm Age}\mid\mathcal{D}_{-i})$ and evaluate the corresponding reweighted CDF at that draw, yielding a PIT-like statistic. By construction, the resulting values are expected to be approximately uniform if the importance weights and the resampling procedure are implemented consistently. We compare the empirical CDF to the uniform reference using the 95\% Dvoretzky–Kiefer–Wolfowitz simultaneous confidence band \citep{Dvoretzky1956}. The resulting ECDF is compatible with uniformity (Fig. \ref{fig:ECDF_DKW}), supporting the internal consistency of the PSIS-LOO reweighting used in the influence analysis.

This result is consistent with the asymmetry towards positive $z_{\rm age}$ values discussed in the main text, which reflects small systematic shifts due to the non-linear LDB structure at the individual level but remains balanced in the global PIT.

\section{PPC metrics on Pleiades}\label{ap:ppc}

\noindent We summarise here quantitative posterior predictive check (PPC) diagnostics derived from replicated datasets. For each observable we compute posterior predictive $z$-scores, predictive interval coverage, and PIT-based statistics. For $A_{\rm Li}$, the predictive $z$-scores are centred close to zero ($\langle z\rangle\simeq0.22$, $\sigma_z\simeq0.97$), but the predictive intervals are systematically wider than expected (e.g. the nominal 50\% interval contains $\sim94\%$ of the observations), and the PIT variance ($\simeq0.015$) is smaller than the uniform expectation. This indicates an over-dispersed posterior predictive distribution. As discussed in the main text, this behaviour originates from the observational layer and the bounded nature of the abundance scale, leading to an accumulation of probability near the Li-rich plateau. For $T_{\rm eff}$, the $z$-score distribution exhibits heavy tails ($\sigma_z\simeq2.8$), driven by a small subset of hot stars near or beyond the validity range of the NN calibration. Within the LDB-sensitive regime, residual behaviour remains well-behaved.

\begin{table}[ht]
\centering
\caption{Posterior predictive check (PPC) metrics.}
\label{tab:ppc_metrics}
\renewcommand{\arraystretch}{1.3}
\begin{tabular}{lcc}
\toprule
\textbf{Metric} & $T_{\rm eff}$ & $A_{\rm Li}$ \\
\hline
Fraction $|z|>2$                                & 0.066 & 0.016 \\
Fraction $|z|>3$                                & 0.049 & 0.016 \\
Coverage (50\%)                                 & 0.85  & 0.94 \\
Coverage (80\%)                                 & 0.91  & 0.96 \\
Coverage (95\%)                                 & 0.93  & 0.96 \\
PIT mean                                        & 0.50  & 0.53 \\
PIT variance                                    & 0.031 & 0.015 \\
Fraction PIT $<0.05$                            & 0.008 & 0.001 \\
Fraction PIT $>0.95$                            & 0.066 & 0.041 \\
\hline\hline
\end{tabular}
\renewcommand{\arraystretch}{1.0}
\tablefoot{Reported quantities include the fraction of sources with large residuals, empirical coverage of posterior predictive intervals, and summary statistics of PIT distributions.}
\end{table}

\section{Probability integral transform (PIT)}\label{ap:PIT}

\noindent Probability integral transform (PIT) diagnostics are fundamentally used to assess predictive calibration by mapping each observation to the interval $[0,1]$. If the predictive model is correctly specified, PIT values follow a uniform distribution. In this work, PIT values are computed under $K$-fold cross-validation and LOO-based procedures to evaluate predictive performance and internal consistency. This appendix summarises their definition and implementation in the context of refitted hierarchical models, and their interpretation under correct predictive calibration. Let $y_i$ denote an observed quantity for star $i$, and let $D_{-i}$ be the dataset used for training, obtained by excluding star $i$ from the fit. Denoting by $p(\Theta \mid D_{-i})$ the posterior distribution of the model parameters given the training data, the LOO predictive cumulative distribution function (CDF) for $y_i$ is defined as
\begin{equation}
F_{-i}(y_i)\equiv F(y_i\mid D_{-i})
=
\int F(y_i\mid\Theta)\,p(\Theta\mid D_{-i})\,\mathrm{d}\Theta,
\end{equation}
where $F(y_i\mid\Theta)$ is the model-implied CDF of the observable conditional on parameters $\Theta$. This quantity represents the fully Bayesian predictive distribution for a new observation drawn under the same data-generating process.

In the context of the hierarchical model, the conditional CDF $F(y_i\mid\Theta)$ is derived from the predictive distribution, $p(y_i\mid\Theta)$, obtained by marginalising over the latent variables, $\boldsymbol{Z}_i$,
\begin{equation}
p(y_i\mid\Theta)
=
\int p(y_i\mid\boldsymbol{Z}_i)\,p(\boldsymbol{Z}_i\mid\Theta)\,\mathrm{d}\boldsymbol{Z}_i
\end{equation}

The corresponding CDF $F(y_i\mid\Theta)$ is then defined as the cumulative distribution associated with this predictive density,
\begin{equation}
F(y_i\mid\Theta)
=
\int_{-\infty}^{y_i} p(y\mid\Theta)\,\mathrm{d}y.
\end{equation}

In practice, the integral above is approximated via Monte Carlo using posterior draws obtained from refitting the model to $D_{-i}$. For each held-out star and each observable, we evaluate the predictive CDF at the observed value $y_{i,\mathrm{obs}}$ to obtain the PIT value given by $\mathrm{PIT}_{i}\;\equiv\; F_{-i}(y_{i,\mathrm{obs}})$. In this work, the PIT diagnostic is applied to effective temperature $T_{\rm eff}$ and lithium abundance $A_{\rm Li}$, for which the observational likelihood is assumed to be Gaussian.

Under this assumption, and using $S$ posterior draws from the refitted model, the PIT can be approximated as
\begin{equation}
\mathrm{PIT}_{i}
\;\approx\;
\frac{1}{S}\sum_{s=1}^{S}
\Phi\!\left(
\frac{y_{i,\mathrm{obs}}-\mu_{i}^{(s)}}{\sigma_{i}^{(s)}}
\right)
,\end{equation}
where $\Phi$ is the standard normal cumulative distribution function, and $(\mu_{i}^{(s)},\sigma_{i}^{(s)})$ are the predictive mean and scale for star $i$ under posterior draw $s$. These quantities incorporate uncertainty from both the hierarchical model parameters and the latent stellar quantities inferred from the training data.

To obtain an out-of-sample diagnostic, we adopt a star-wise $K$-fold cross-validation scheme. The dataset is partitioned into $K$ subsets, and for each fold the model is refitted after removing all observables of the held-out stars. A single PIT value is then computed per star (and observable) using the predictive distribution from the corresponding refit. This ensures that all PIT values are evaluated out-of-sample. The resulting PIT distributions are summarised using histograms and empirical cumulative distribution functions (ECDFs), with Dvoretzky–Kiefer–Wolfowitz (DKW) confidence bands (Appendix \ref{ap:dkw}) providing a reference for finite-sample deviations from uniformity. For directly observed quantities such as $T_{\rm eff}$ and $A_{\rm Li}$, PIT uniformity provides a meaningful test of predictive calibration. For global latent parameters (e.g. the LDB age), PIT-like diagnostics are instead interpreted as checks of internal consistency rather than formal calibration tests.

\section{Posterior probability of rotational membership}
\label{ap:prob_rot}

\noindent Within the hierarchical formulation described in Sect. \ref{sc:bhm}, rotational effects are incorporated through a latent two-component mixture for the true lithium abundance $A_{{\rm Li,true},i}$ of each star $i$ (Eq. \ref{eq:latent_two_component_mixture}). In this model, each source is assumed to originate either from a quiet (or slow rotators) population or from a population of fast rotating stars, with prior mixing weight $\omega_i$ given by Eq. (\ref{eq:prob_i}). In this equation, we denote by $\omega_{\mathrm{FGK},i}$ the probability that star $i$ belongs to the FGK regime and $p_{\rm rot}$ is the global fraction of fast rotators within that regime. Although the discrete indicator variable specifying the population membership of each star is marginalised in the model, the posterior distribution naturally allows us to quantify the probability that a given star belongs to the fast rotators component. This quantity is used as a diagnostic and visualisation tool in Fig. \ref{fig:LDD_z}. For each source $i$, we define the posterior probability of fast rotators membership as
\begin{equation}
P_i^{\,\rm rot}
\;\equiv\;
\mathbb{E}_{p(\Theta,\boldsymbol{Z}\mid\boldsymbol{D})}
\!\left[
P\!\left(
z_i=\mathrm{rot}
\;\middle|\;
A_{\mathrm{Li,true},i},\Theta
\right)
\right]
\label{eq:prot_def}
,\end{equation}
where $z_i\in\{\mathrm{quiet},\mathrm{rot}\}$ denotes the latent mixture component for star $i$, $\Theta$ the set of global parameters, and $A_{\mathrm{Li,true},i}$ the latent true lithium abundance.

The inner conditional probability in Eq. (\ref{eq:prot_def}) corresponds to the standard mixture responsibility for the fast rotators component, evaluated in latent space,
\begin{equation}
 P\!\left(
z_i=\mathrm{rot}
\;\middle|\;
A_{\mathrm{Li,true},i},\Theta
\right) =
\frac{1}{\mathcal{C}}
\omega_i\,
f_{\rm rot}\!\left(A_{\mathrm{Li,true},i}\mid\Theta\right)
\label{eq:responsibility},
\end{equation}
where $\mathcal{C}$ is defined as the mixture normalisation constant:
\begin{equation}
\mathcal{C} = (1-\omega_i)\,
f_{\rm quiet}\!\left(A_{\mathrm{Li,true},i}\mid\Theta\right)
+
\omega_i\,
f_{\rm rot}\!\left(A_{\mathrm{Li,true},i}\mid\Theta\right),
\end{equation}

The component densities entering Eq. (\ref{eq:responsibility}) are given by
\begin{equation}
\begin{split}
& f_{\rm quiet}
=
\mathcal{N}\!\left(
\mu_{A_{\mathrm{Li}},i},\,
\sigma_{A_{\rm Li}}^{2}
\right),\\
& f_{\rm rot}
=
\mathcal{SN}\!\left(
\mu_{A_{\mathrm{Li}},i}
+
\Delta A_{{\rm Li},\rm rot},\,
\sigma_{A_{\rm Li}}^{2}
+
\sigma_{A_{\rm Li},\rm rot}^{2},\,
\alpha_{\rm FGK}
\right),
\end{split}
\end{equation}
where $\mathcal{N}$ denotes a Normal distribution and $\mathcal{SN}$ a Skew-Normal distribution with shape parameter $\alpha_{\rm FGK}$.

In practice, Eq. (\ref{eq:prot_def}) is evaluated using posterior samples obtained from the Markov chain Monte Carlo inference. For each posterior draw $(\Theta{(s)},\boldsymbol{Z}{(s)})$, the responsibility in Eq. (\ref{eq:responsibility}) is computed using the sampled value of $A_{\mathrm{Li,true},i}{(s)}$ and the corresponding global parameters. The final probability $P_i^{\,\rm rot}$ is then obtained by averaging these responsibilities over all posterior samples. This definition yields a continuous probability in the interval $[0,1]$, which reflects both the uncertainty in the latent lithium abundance and the uncertainty in the global model parameters. Values close to $0$ indicate stars that are consistently explained by the quiet component, while values close to $1$ correspond to stars that are robustly associated with the fast rotating population. This quantity is therefore well suited for diagnostic plots and for assessing the role of rotation in shaping the lithium distribution near the LDB, without introducing any additional assumptions beyond those already encoded in the model.

We further verified the robustness of this probabilistic classification by excluding sources independently identified as fast rotators based on their projected rotational velocities. The resulting membership probabilities remain correlated with rotation classification, as lithium preservation at temperatures where depletion should already have occurred is unlikely in the absence of rotational effects. Therefore, the model consistently attributes enhanced $A_{\rm Li}$ to the fast rotation population.

\section{Dvoretzky–Kiefer–Wolfowitz inequality}\label{ap:dkw}

\noindent In several parts of this work, empirical cumulative distribution functions (ECDFs) of probability integral transform (PIT) values are used as diagnostics of predictive calibration and internal consistency under cross-validation. In particular, when assessing the uniformity of PIT values obtained either from $K$-fold refitting or from LOO-reweighted posterior predictive checks, it is necessary to quantify the level of deviation from the uniform distribution that can be attributed to finite-sample variability alone. The Dvoretzky–Kiefer–Wolfowitz (DKW) inequality provides a non-asymptotic, distribution-free bound that enables the construction of simultaneous confidence bands for the ECDF, which we adopted throughout this work.

Let $\widehat{F}_N(x)$ denote the empirical cumulative distribution function of $N$ PIT values,
\begin{equation}
\widehat{F}_N(x) \;=\; \frac{1}{N}\sum_{i=1}^N \mathbf{1}\{\mathrm{PIT}_i \le x\},
\end{equation}
and let $F_0(x)$ be the reference cumulative distribution function under the null hypothesis of correct calibration. For PIT diagnostics, this null corresponds to $F_0(x) = x$, $x\in[0,1]$, i.e. a $\mathcal{U}(0,1)$ distribution.

The Dvoretzky–Kiefer–Wolfowitz inequality states that, for any $\varepsilon>0$, we then have
\begin{equation}
\Pr\left(
\sup_{x}\left|\widehat{F}_N(x)-F_0(x)\right|>\varepsilon
\right)
\;\le\;
2\,e^{-2N\varepsilon^2},
\end{equation}

This bound controls the probability that the maximum absolute deviation between the ECDF and the reference CDF exceeds a given threshold $\varepsilon$, uniformly over the entire domain of $x$. Importantly, the inequality is non-asymptotic and holds for any sample size, $N$, without assumptions on the form of $F_0$, apart from independent and identically distributed samples.

By setting the right-hand side equal to a chosen significance level $\alpha\in(0,1)$, one obtains the half-width of a simultaneous confidence band,
\begin{equation}\label{eq:epsilon}
\varepsilon(\alpha,N)
\;=\;
\sqrt{\frac{\ln(2/\alpha)}{2N}}.
\end{equation}

Then, with probability at least $1-\alpha$, the entire ECDF lies within the envelope, expressed via  $F_0(x)-\varepsilon(\alpha,N)\le \widehat{F}_N(x) \le F_0(x)+\varepsilon(\alpha,N), \: \forall x$.

In practice, since both $F_0(x)$ and $\widehat{F}_N(x)$ take values in $[0,1]$, the bounds are clipped to this interval. For example, the width of the DKW band from Fig. \ref{fig:kfold_pit_ecdf}, $\varepsilon\simeq0.12$, corresponds to a 95\% confidence level ($\alpha=0.05$) for a sample size $N=122$, computed from Eq. (\ref{eq:epsilon}).

In the context of this work, these DKW bands are overlaid on ECDFs of PIT values to provide a visual and quantitative reference for expected sampling variability under correct calibration. Because the bands are simultaneous in $x$, they control global deviations of the ECDF rather than pointwise fluctuations. Consequently, systematic departures of the ECDF outside the DKW envelope indicate deviations from uniformity that cannot be attributed to finite-sample noise at the chosen confidence level. We note that, in cases where PIT values are constructed from global latent parameters (e.g. the LDB age under LOO reweighting), the DKW-based check should be interpreted as a diagnostic of internal consistency of the reweighting and resampling procedure, rather than as a formal test of predictive calibration. For observable quantities evaluated under $K$-fold cross-validation, however, the same DKW bands provide a meaningful and interpretable criterion for assessing predictive calibration of the hierarchical model.

\end{appendix}

\end{document}